\documentclass[12pt]{article}

\pdfoutput=1

\usepackage[T1]{fontenc}
\usepackage[utf8]{inputenc}
\usepackage[english]{babel}
\usepackage{amsmath,amssymb,amsfonts,mathtools}
\usepackage{amsthm}
\usepackage{latexsym}
\usepackage[mathcal]{euscript}
\usepackage{graphicx}
\usepackage{epsfig}
\usepackage{subfig}
\usepackage{psfrag}
\usepackage{array}
\usepackage{float}
\usepackage{cancel}
\usepackage{mathrsfs}
\usepackage{bm}
\usepackage[normalem]{ulem}
\usepackage{xcolor}
\usepackage{slashed}
\usepackage{appendix}
\usepackage{jheppub}

\newcommand{\be}{\begin{equation}}
\newcommand{\ee}{\end{equation}}
\newcommand{\bea}{\begin{eqnarray}}
\newcommand{\eea}{\end{eqnarray}}
\newcommand{\bi}{\begin{itemize}}
\newcommand{\ei}{\end{itemize}}

\definecolor{dgreen}{rgb}{0.,0.6,0.}

\begin{document}

\title{The Schrodinger Equation as a Gauge Theory}

\author[a,b]{Dmitry S. Ageev}
\author[b,a]{and Vladimir A. Bykov}

\affiliation[a]{Department of Mathematical Methods for Quantum Technologies, Steklov Mathematical Institute of Russian Academy of Sciences, Gubkina str. 8, Moscow 119991, Russia}
\affiliation[b]{Institute for Theoretical and Mathematical Physics, Lomonosov Moscow State University, 119991 Moscow, Russia}

\emailAdd{ageev@mi-ras.ru}
\emailAdd{bykov.va20@physics.msu.ru}

\abstract{
The conserved probability current of the Schrodinger field admits a dimension-dependent gauge representation that provides a convenient setting for comparing hydrodynamic, topological and infrared structures. We use this representation to develop a unified treatment of several deformations of the Madelung momentum. BF couplings incorporate electromagnetic interactions, Berry and spin connections, projected non-abelian adiabatic data, and intrinsic phase holonomy within the same framework. In $(2+1)$ dimensions, the Chern-Simons deformation can be reduced to a nonlocal functional of the wavefunction. Its Coulomb-gauge form separates into a phase-sensitive geometric contribution and a density-dependent dynamical contribution; we show explicitly how this separation changes under a general gauge transformation and recover the topological braiding phase in the well-separated adiabatic regime. For systems with a spatial boundary, we derive the equal-time symplectic structure, the associated surface charges, and their boundary algebra. Finally, after a nonlinear interaction produces the Bogoliubov acoustic branch, the radiative memory can be expressed in the dual two-form variables. Its zero-average component is generated by a residual large gauge transformation, whereas the monopole component remains independent global data. This provides a gauge-theoretic description of the memory/asymptotic-symmetry sector of the acoustic infrared structure.
}

\maketitle

\section{Introduction}

It is often a useful moment in field theory when an equation that first appeared as dynamics turns out, after the right change of variables, to be kinematics. Recent work on fluids has repeatedly exhibited precisely this phenomenon. Once the continuity equation is rewritten in the language of differential forms, the conserved density-current sector can be packaged into gauge fields, and structures that ordinarily enter only later---vorticity, topological couplings, boundary charges, memory observables---move to the foreground from the start. The recent gauge-theoretic formulations of shallow water \cite{Tong:2022gpg}, incompressible Euler flow, perfect fluids in terms of Kalb--Ramond fields, and more general fluid/$p$-form systems have made this lesson especially clear \cite{Nastase:2023rou,Eling:2023iyx,Eling:2023apf,Matsuo:2021xas,Taghiloo:2023mtg}. The lesson itself is simple, but it also extends beyond any one example, as conservation laws often signal that a theory wants to be rewritten in gauge language. Related group-theoretic and coadjoint-orbit formulations of fluid dynamics have been developed for non-abelian and perfect fluids by Jackiw, Nair and collaborators \cite{Bistrovic:2002jx,Jackiw:2004nm}. In this framework, non-abelian fluids can also support topologically nontrivial solitonic configurations, including the color skyrmions of \cite{Dai:2006te}. Related constructions incorporate anomalous transport and spin degrees of freedom \cite{Nair:2011mk,Karabali:2014vla}, while topological terms and their role in fluid dynamics were studied in \cite{Nair:2020kjg}.

The starting point of this paper is that the nonrelativistic Schrodinger equation is much closer to this pattern than one might first think. The wavefunction carries a density and a conserved probability current. Away from its nodal set it also defines a phase and therefore a velocity potential. In the usual Madelung representation the imaginary part of the Schrodinger equation becomes the continuity equation, while the real part becomes a Hamilton--Jacobi equation corrected by the quantum potential. In that form the theory already looks like a compressible fluid, and this viewpoint is classical and well established, going back to Madelung and developed systematically, especially for spinful systems, by Takabayasi and others \cite{Takabayasi:1952, Takabayasi:1953, Takabayasi:1954, Takabayasi:1955, Takabayasi:1956Dirac, Takabayasi:1956Variational, Takabayasi:1957Dirac, TakabayasiVigier:1957, Takabayasi:1983zh, Bialynicki-Birula:1995fal, Recami:1995iy, Trukhanova:2018gmh, Trukhanova:2021qbe, Trukhanova:2021smo, Fabbri:2022kfr, Fabbri:2023onb, Fabbri:2023yhl, Fabbri:2025ftm, DAlessi:2023zop}. Closely related density--phase hydrodynamic formulations of interacting Bose
systems were developed by Popov \cite{Popov:1972hyd}. The hydrodynamic presentation, however, leaves current conservation as one of the equations of motion. Here we instead take it as the entry point to a gauge description.

The gauge structure then follows from the conserved probability current. This step is itself standard: the Hodge dual of a conserved current is closed and is therefore locally exact. In $(2+1)$ dimensions this gives the familiar current--gauge correspondence underlying particle--vortex duality \cite{Peskin:1977kp,Dasgupta:1981zz}, while in $(3+1)$ dimensions the current is naturally represented by the field strength of an abelian two-form gauge field. The latter structure appears in the Kalb--Ramond description of perfect fluids and in the general fluid/$p$-form dictionary \cite{Matsuo:2021xas,Taghiloo:2023mtg}. In either case the continuity equation becomes a Bianchi identity. What remains as dynamics are the conditions that, in hydrodynamic language, express irrotational flow and the quantum Euler equation. This correspondence is local. The gauge and fluid variables encode only derivatives of the phase, while the original wavefunction imposes the additional requirement of single-valuedness. Around nodes of $\psi$ the missing information reappears as quantized circulation. Thus the Schrodinger equation, quantum hydrodynamics, and the gauge theory are three local presentations of the same system, with global sectors labeled by the phase winding data.

The current--gauge correspondence places the Schrodinger system within a broader class of gauge and geometric formulations of fluid dynamics. In shallow water theory, the gauge formulation identifies the linearized Poincare waves with Maxwell-Chern-Simons modes and the coastal Kelvin waves with chiral edge modes~\cite{Tong:2022gpg}, while the topological origin of equatorial Kelvin and Yanai waves was clarified from the bulk topology of the Poincare bands \cite{Delplace:2017izb}. For Euler flow, the gauge formulation gives a topological argument for the quantization of the Euler Hopfion \cite{Nastase:2023rou}. In related fluid and topological systems, gauge variables have also been used to analyze boundary algebras, Chern-Simons and BF edge modes \cite{Eling:2023iyx, Eling:2023apf, Sheikh-Jabbari:2023eba, Bertolini:2026qit}, fractional and integer quantum Hall hydrodynamics \cite{Monteiro:2022wip, Reynolds:2024vqn}, topological waves with odd viscosity \cite{Fujii:2024bdb}, superfluid waves and charge-vortex duality \cite{Hsiao:2024rbg, Tsaloukidis:2025jzj}.

Within this framework, electromagnetic coupling, spin and the holonomy one-form attached to the phase of the wavefunction \cite{Tronci:2020vlq} all modify the phase-gradient sector. In the gauge description these modifications are implemented uniformly by a BF coupling to an additional one-form, which shifts the Madelung momentum on the hydrodynamic side. For example, for a two-component wavefunction this one-form is the Berry connection on the Hopf bundle. In a degenerate adiabatic sector the Berry connection becomes non-abelian \cite{Mead:1979ayo, Wilczek:1984dh, Ruseckas:2005vgb, Dalibard:2010ph}, and BF structure applies after projecting it onto the occupied polarization, which selects the abelian one-form seen by the hydrodynamic current.

The Chern-Simons term gives the complementary topological deformation. Chern-Simons-coupled nonrelativistic bosonic fluids are well established, most notably in the composite-boson description of the fractional quantum Hall effect \cite{Zhang:1988wy}, as well as in Chern-Simons Schrodinger systems such as \cite{Jackiw:1990tz}. Their relation to charge--flux attachment, fractional statistics and Hopf interactions is also well known \cite{Wilczek:1983cy,Wu:1984kd}. Here we use this structure within the present current--gauge dictionary to trace the deformation through the fluid and wavefunction variables. 

The presence of a boundary provides a further setting in which the gauge structure becomes nontrivial. Related boundary structures have been studied in topological-fluid descriptions of fractional quantum Hall edge dynamics, where the chiral boson action can be obtained directly from the fluid formulation \cite{Monteiro:2022wip}. Residual gauge transformations can act nontrivially on the boundary and give rise to physical edge degrees of freedom. We determine boundary conditions for a well-posed variational principle, construct the associated surface charges, and derive their equal-time algebra for the Chern-Simons and spin/BF deformations.

The gauge/fluid viewpoint also opens a natural way to discuss the long-distance sector of the theory. Memory effects have recently been studied in shallow water theory \cite{Sheikh-Jabbari:2023eba}, acoustic sound \cite{deAguiarAlves:2025vfu}, Bose--Einstein condensates \cite{Datta:2020rrf}, and superfluids \cite{Jensen:2026hrx}. This motivates asking whether an analogous acoustic memory effect can arise in the Schrodinger system and, if so, how it is encoded in its dual two-form gauge description.

For the linear Schrodinger equation, however, the infrared excitations are too dispersive to support an ordinary acoustic memory effect, since linearization around a homogeneous background gives $\omega\sim k^2$ rather than a sound mode. Once a local nonlinear interaction is added, the Bogoliubov spectrum contains a genuine acoustic branch. In this regime the $(3+1)$-dimensional two-form gauge description provides a natural description of the infrared sector. The acoustic memory follows from the retarded response of the Bogoliubov mode. We then express the same memory in the dual two-form variables and show that its zero-average part is generated by a residual large gauge transformation, while its monopole component remains an independent global mode.

The relation between memory effects, asymptotic symmetries, and soft theorems is the infrared triangle familiar from gauge theory and gravity \cite{Weinberg:1965nx,Strominger:2013jfa,He:2014cra,Strominger:2014pwa,Tolish:2014bka,Satishchandran:2019pyc,Campiglia:2015qka,Bieri:2013hqa,Pasterski:2015zua}, with its gravitational roots in displacement memory and BMS symmetry \cite{Bondi:1962px,Sachs:1962wk,Zeldovich:1974gvh}. The infrared-triangle picture has also been extended beyond its conventional relativistic setting, notably to hydrodynamic sound \cite{deAguiarAlves:2025vfu} and to fracton theories \cite{Perez:2023uwt}. Here the nonlinear Schrodinger system realizes the memory and asymptotic-symmetry sides of this structure in the acoustic regime, while the corresponding soft sector is expected to provide the third corner.

The contribution of the present work is to use the gauge formulation as a unifying language for structures that otherwise appear separately in the Schrodinger, hydrodynamic and topological descriptions. BF couplings provide a systematic dictionary for electromagnetic, Berry, spin and holonomy effects, while in the Chern-Simons deformation the wavefunction functional separates in Coulomb gauge into geometric and density-dependent contributions, and we determine how this decomposition changes under a general gauge transformation. With boundaries, the same framework makes it possible to identify the admissible edge degrees of freedom, construct their surface charges and determine the corresponding equal-time algebras. In the nonlinear $(3+1)$-dimensional theory, the dual two-form description also gives a gauge-theoretic packaging of acoustic memory and relates its zero-average part directly to a residual large gauge transformation.

The paper is organized as follows. In Section~\ref{sec:schr-qhydro-gauge} we derive the local correspondence between the Schrodinger equation, quantum hydrodynamics, and gauge theory, and separate this local equivalence from the global phase data carried by the wavefunction. In Section~\ref{sec:topological-terms} we study topological deformations, including BF couplings, Chern-Simons terms, spin extensions, and intrinsic holonomy. In Section~\ref{sec:edge-modes} we place the theory on space with a boundary and derive the corresponding edge charges and boundary algebras. In Section~\ref{sec:memory} we turn to the nonlinear infrared regime and show how sound, memory, and asymptotic symmetry are organized by the dual two-form description. Technical details are collected in Appendices~\ref{app:notation}--\ref{app:derivation}.

\section{The Schrodinger/Fluid/Gauge correspondence}
\label{sec:schr-qhydro-gauge}

The starting point of the correspondence is that the wavefunction already contains within it the variables of a compressible quantum fluid, while the corresponding conserved current admits a gauge-theoretic representation. It is useful to view these facts as parts of a chain of locally equivalent descriptions. Once the continuity equation is rewritten in a form solved identically, the Schrodinger system acquires a gauge-theoretic formulation. We will work throughout in $d=2$ and $d=3$ spatial dimensions. The hydrodynamic description is common to both cases, while the gauge-theoretic packaging of the conserved current depends on the dimension.

\subsection*{From the wavefunction to the quantum fluid}

We begin with the Schrodinger equation with an arbitrary external potential
\be
i\hbar \partial_t \psi
=
-\frac{\hbar^2}{2m}\nabla^2 \psi
+
V(\mathbf x,t)\psi.
\label{Schrodinger eq}
\ee
On any patch where the wavefunction does not vanish, we can write it in Madelung form
\be
\psi(\mathbf x,t)
=
\sqrt{\rho(\mathbf x,t)}\,
e^{iS(\mathbf x,t)/\hbar}.
\label{Madelung trans}
\ee
Substituting the decomposed wavefunction \eqref{Madelung trans} into the equation \eqref{Schrodinger eq} and separating the imaginary and the real parts gives two equations. The imaginary part is the continuity equation
\be
\partial_t \rho+\nabla\cdot(\rho \mathbf v)=0,
\label{eq: continuity}
\ee
with velocity field
\be
\mathbf v\equiv \frac{1}{m}\nabla S,
\label{eq:v=nabla S}
\ee
defined away from the nodal set of \(\psi\). This is hydrodynamic form of probability conservation equation with 
\begin{gather}
\rho=\psi^*\psi,\\
\rho \mathbf v=-\frac{i\hbar}{2m}(\psi^*\nabla \psi -\nabla \psi^*  \psi).
\end{gather}
The real part gives the quantum Hamilton--Jacobi equation,
\be
\partial_t S+\frac{(\nabla S)^2}{2m}+V+Q=0,
\qquad
Q\equiv -\frac{\hbar^2}{2m}\frac{\nabla^2\sqrt{\rho}}{\sqrt{\rho}},
\label{eq: quantum Euler}
\ee
where the term $Q$ is the quantum potential. The hydrodynamic interpretation of \eqref{eq: quantum Euler} becomes transparent after taking a spatial gradient. In \(d=2\) the vorticity is the scalar
\be
\omega\equiv \varepsilon^{ij}\partial_i v_j,
\ee
while in \(d=3\) it is the vector
\be
\boldsymbol{\omega}\equiv \nabla\times \mathbf v.
\ee
Because \(\mathbf v\) is locally a gradient, the flow is automatically irrotational wherever \(S\) is smooth and \(\psi\neq 0\). The only place where vorticity can reside is therefore at singularities of the phase, namely at nodes of the wavefunction where \(\rho=0\) and the Madelung variables cease to be regular. Using
\begin{gather}
\partial_k\!\left(\frac12 v^2\right)
=
(\mathbf v\cdot\nabla) v_k+\omega\,\varepsilon_{kj}v_j,
\qquad d=2,
\\[2mm]
\nabla\!\left(\frac12 v^2\right)
=
(\mathbf v\cdot\nabla)\mathbf v+\mathbf v\times \boldsymbol{\omega},
\qquad d=3,
\end{gather}
and then setting the vorticity to zero away from nodes, one finds in both dimensions
\be
m\left(\partial_t+\mathbf v\cdot\nabla\right)\mathbf v
=
-\nabla(V+Q).
\label{eq: Euler eq}
\ee
Equations \eqref{eq: continuity} and \eqref{eq: Euler eq} are the equations of an inviscid compressible quantum fluid. The essential difference from ordinary Euler hydrodynamics is the force \(-\nabla Q\), which has no classical analogue. In particular, \(\nabla Q\) contains third spatial derivatives of \(\rho\), so once the theory is written in fluid variables the dynamics becomes higher-derivative in space even though the original Schrodinger equation is only second order.

\subsection*{Global phase data}

At this stage one might be tempted to identify \((\rho,\mathbf v)\) with the full content of the wavefunction. That statement is correct locally, but not globally. The wavefunction $\psi$ contains information about the phase \(S\), whereas the fluid only knows its derivative. As a result, the phase can be reconstructed only locally, and only up to an additive function of time. If one treats \((\rho,\mathbf v)\) as unconstrained fields, one enlarges the space of configurations by allowing velocity profiles that do not come from a single-valued wavefunction.

The missing global data is the set of periods of the one-form \(dS\). Single-valuedness of \(\psi\) requires that the phase change along any closed contour \(C\) be an integer multiple of \(2\pi\hbar\), which is equivalent to the circulation quantization condition
\be
\oint_C \mathbf v\cdot d\mathbf l
=
\frac{2\pi\hbar}{m}\,n,
\qquad
n\in \mathbb Z.
\label{eq:qunt cond}
\ee
For contours that do not enclose the nodal set one has \(n=0\), in agreement with local irrotationality. Nonzero circulation is possible only when the contour winds around zeros of \(\psi\).

\subsection*{Variational principle for quantum fluid}

For the following gauge formulation it is useful to record the action that reproduces this hydrodynamic system. Starting from the Schrodinger action and substituting the Madelung form (\ref{Madelung trans}), one finds, up to a total derivative
\be
I[\rho,S]
=
\int dt\, d^dx\,
\left[
-\rho\,\partial_t S
-\frac{\rho(\nabla S)^2}{2m}
-U(\rho,\nabla\rho)
\right],
\label{eq:Madelung action}
\ee
with
\be
U(\rho,\nabla\rho)
=
V\rho+\frac{\hbar^2}{8m}\frac{(\nabla\rho)^2}{\rho}.
\ee
An equivalent first-order form is obtained by introducing the current $\mathbf j=\rho\nabla S/m$ as an auxiliary field
\be
I[\rho,\mathbf j,S]
=
\int dt\, d^dx\,
\left[
-\rho\,\partial_t S
-\mathbf j\cdot \nabla S
+\frac{m\mathbf j^2}{2\rho}
-U(\rho,\nabla\rho)
\right].
\ee
After integrating by parts, this becomes
\be
I[\rho,\mathbf j,S]
=
\int dt\, d^dx\,
\left[
S\big(\partial_t\rho+\nabla\cdot \mathbf j\big)
+\frac{m\mathbf j^2}{2\rho}
-U(\rho,\nabla\rho)
\right],
\label{eq:first-order-fluid-action}
\ee
so \(S\) appears as a Lagrange multiplier imposing the continuity equation. This first-order formulation is the starting point for the gauge description, since the gauge fields introduced below only solve the continuity constraint off-shell.

\begingroup

The continuity equation is the statement that the current $J^\mu=(\rho,j^i)$ is conserved, $\partial_\mu J^\mu=0$. Locally its Hodge dual is therefore exact, so the continuity equation may be solved in terms of gauge potentials. In $(2+1)$ dimensions this is the continuum current--gauge map familiar from particle--vortex duality \cite{Peskin:1977kp,Dasgupta:1981zz}; in $(3+1)$ dimensions the dual current is a closed three-form and is represented by the field strength of a two-form gauge potential, in accordance with the general fluid/$p$-form dictionary \cite{Taghiloo:2023mtg}.
\endgroup

\subsection*{Gauge formulation in $(2+1)$ dimensions}

We now restrict attention to two spatial dimensions. Our notation is as follows, and is summarized in Appendix~\ref{app:notation}. Greek indices label spacetime components, while Latin indices label Euclidean spatial components only. We take $
x^\mu=(t,x^i)$, with $i=1,2$,
and use the antisymmetric symbols $\varepsilon^{\mu\nu\rho}$ with covariant indices and $\varepsilon^{ij}=\varepsilon^{0ij}$ with Euclidean: $\varepsilon^{012}=+1$, $\varepsilon^{12}=+1$.
Spatial indices are raised and lowered with \(\delta_{ij}\). For the gauge potential it is important to distinguish spacetime one-form components from Euclidean vector components: we write $
A_\mu=(A_0,-A_i)$, 
$A^\mu=(A_0,A^i)$,
with \(A^i=\delta^{ij}A_j\). Thus the minus sign belongs to the spacetime one-form convention, whereas \(A_i\) and \(A^i\) are related only by the Euclidean metric. 

In these conventions the gauge-invariant field strengths are
\be
B\equiv \varepsilon^{ij}\partial_i A_j,
\qquad
E_i\equiv -\partial_t A_i-\partial_i A_0.
\ee
The field strength furnishes an identically conserved current. Indeed, if we define
\be
2J^\mu=-\varepsilon^{\mu\alpha\beta}F_{\alpha\beta},
\qquad
F_{\alpha\beta}=\partial_\alpha A_\beta-\partial_\beta A_\alpha,
\ee
then the Bianchi identity for \(F\) implies \(\partial_\mu J^\mu=0\) identically. We therefore identify the hydrodynamic density and current as
\be
\rho\equiv B\equiv J^0,
\qquad
j^i\equiv \rho v^i\equiv \varepsilon^{ij}E_j\equiv J^i.
\label{eq:hydro-gauge-ident-2d}
\ee
With this identification, the continuity equation is no longer a dynamical equation but a kinematic identity,
\be
\partial_t B+\varepsilon^{ij}\partial_iE_j=0
\qquad\Longleftrightarrow\qquad
\partial_t\rho+\partial_i j^i=0.
\ee
Since $\rho\geq0$, the physical sector satisfies $B\geq0$. On the
regular patches considered below, away from the nodal set, we restrict
to $B>0$.

From the viewpoint of the first-order fluid action \eqref{eq:first-order-fluid-action}, this means that the continuity constraint has now been solved explicitly. Substituting \(\rho=B\) and \(j^i=\varepsilon^{ij}E_j\) gives
\be
I[A_\mu]
=
\int dt\, d^2x\,
\left(
\frac{m\mathbf E^2}{2B}
-U(B,\nabla B)
\right),
\label{eq:gauge action for Schr (1+2)}
\ee
where the first term is the hydrodynamic energy density rewritten in gauge variables. The denominator \(B=\rho\) also makes clear that the gauge description is well defined precisely on the same patches where the Madelung variables are regular, while the singular behavior at \(B=0\) is the gauge-theory reflection of the nodal set.

The remaining equations of motion reproduce the irrotational and Euler sectors of the quantum fluid. Varying with respect to \(A_0\) gives
\be
\partial_i\!\left(\frac{mE_i}{B}\right)=0.
\ee
Since \(\rho=B\) and \(v^i=\varepsilon^{ij}E_j/B\), this is exactly the irrotationality condition written in gauge variables,
\be
\omega\equiv \varepsilon^{ij}\partial_i v_j
=
-\partial_i\!\left(\frac{E_i}{B}\right)=0.
\label{eq:irrot-2d-gauge}
\ee
Varying with respect to \(A_i\) gives the dynamical equation
\be
\partial_t\left(\frac{mE_i}{B}\right)
+
\varepsilon^{ji}\partial_j
\left(
\frac{m\mathbf E^2}{2B^2}+V+Q
\right)
=0,
\qquad
Q=-\frac{\hbar^2}{2m}\frac{\nabla^2\sqrt{B}}{\sqrt{B}}.
\ee
After contracting with \(\varepsilon^{\ell i}\) and using \(v^\ell=\varepsilon^{\ell i}E_i/B\), one obtains
\be
\partial_t(mv_\ell)+\partial_\ell\!\left(\frac12 mv^2+V+Q\right)=0.
\ee
This is the gradient form of the quantum Euler equation, and once the irrotationality condition \eqref{eq:irrot-2d-gauge} is imposed it is equivalent to \eqref{eq: Euler eq}. Thus in \((2+1)\) dimensions the one-form gauge field automatically encodes the density and current through a Bianchi identity, while its equations of motion enforce the remaining hydrodynamic constraints.

\subsection*{Gauge formulation in $(3+1)$ dimensions}

In three spatial dimensions, the dual description is organized in terms of an antisymmetric two-form \(B_{\mu\nu}=-B_{\nu\mu}\), transforming as
\be
B_{\mu\nu}\to B_{\mu\nu}+\partial_\mu \varepsilon_\nu-\partial_\nu \varepsilon_\mu.
\ee
Its gauge-invariant content is captured by the three-form field strength
\be
H_{\mu\nu\rho}
=
\partial_\mu B_{\nu\rho}
+\partial_\nu B_{\rho\mu}
+\partial_\rho B_{\mu\nu},
\ee
whose Bianchi identity reads
\be
\partial_\mu\!\left(\varepsilon^{\mu\nu\alpha\beta}H_{\nu\alpha\beta}\right)=0.
\ee
To connect this description with hydrodynamics, we introduce the combinations
\be
H\equiv \varepsilon^{ijk}H_{ijk},
\qquad
H^i\equiv -3\,\varepsilon^{ijk}H_{0jk}.
\ee
In terms of these variables, the Bianchi identity becomes
\be
\partial_t H+\partial_i H^i=0,
\ee
hence it takes precisely the form of the continuity equation. We are therefore led to identify
\be
\rho\equiv H,
\qquad
\rho v^i\equiv H^i.
\label{eq:H-hydro-ident}
\ee
Likewise, $H=\rho\geq0$, and on a regular patch away from nodes we work with $H>0$.

Substituting this parametrization into the first-order fluid action gives
\be
I[B_{\mu\nu}]
=
\int dt\, d^3x\,
\left(
\frac{mH_iH^i}{2H}
-U(H,\nabla H)
\right).
\ee
Varying with respect to \(B_{0i}\) changes only \(H^i\), with
\be
\delta_{B_{0k}}H=0,
\qquad
\delta_{B_{0k}}H^i
=
6\,\varepsilon^{ijk}\partial_j\delta B_{0k}.
\ee
The resulting equation of motion is
\be
\varepsilon^{ijk}\partial_j\!\left(\frac{mH_i}{H}\right)=0,
\ee
or, using \(v_i=H_i/H\),
\be
\nabla\times \mathbf v=0.
\ee
Thus the temporal components enforce irrotational flow.

Varying with respect to \(B_{jk}\) gives
\be
\delta_{B_{jk}}H
=
3\,\varepsilon^{ijk}\partial_i\delta B_{jk},
\qquad
\delta_{B_{jk}}H^i
=
-3\,\varepsilon^{ijk}\partial_t\delta B_{jk},
\ee
and the equation of motion becomes
\be
\partial_t\!\left(\frac{mH_i}{H}\right)
+
\partial_i\!\left(
\frac{mH_jH^j}{2H^2}+V+Q
\right)
=0.
\ee
Since \(H_i/H=v_i\), this is
\be
\partial_t(mv_i)+\partial_i\!\left(\frac12 mv^2+V+Q\right)=0.
\ee
Together with \(\nabla\times \mathbf v=0\), this reproduces the Euler equation \eqref{eq: Euler eq}.

\begingroup

\subsection*{Time-dependent nodes and nodal topology}

The correspondence between the Schrodinger equation, quantum hydrodynamics and the gauge theory is local. It applies in regions where the wavefunction does not vanish, so that $B=\rho>0$ in $(2+1)$ dimensions and $H=\rho>0$ in $(3+1)$ dimensions. At the nodes, the reconstruction of the velocity and phase becomes singular.

For a time-dependent wavefunction, the nodal set may move without changing its topology. To make this distinction more precise, consider first $(2+1)$ dimensions and write $\psi=u+iv$. An isolated node is regular when
\be
u=v=0,
\qquad
J\equiv\varepsilon^{ij}\partial_i u\,\partial_j v\neq0.
\label{eq:regular-node-2d}
\ee
The implicit-function theorem then implies that the position of the node varies smoothly with time. For a simple zero, the winding number is determined by the sign of $J$ and therefore remains fixed as long as the zero remains regular. The closed contours used to define circulation can consequently be transported continuously with the moving nodes, provided they remain in the region where $\psi\neq0$. The circulation data can then be used for phase reconstruction throughout such an evolution.

A change of nodal topology requires the regularity condition \eqref{eq:regular-node-2d} to fail. This can be seen already for the free Schrodinger equation. Consider the solution
\be
\psi(x,y,t)
=
\psi_0\left(
\frac{x^2}{L^2}
+
(1+i)\frac{y}{L}
+
i\frac{\hbar t}{mL^2}
\right),
\label{eq:node-pair-example}
\ee
where $\psi_0$ is a nonzero constant amplitude and $L>0$ is an arbitrary length scale. Introducing
\be
\tau\equiv\frac{\hbar t}{mL^2},
\label{eq:node-example-tau}
\ee
the conditions $\operatorname{Re}\psi=\operatorname{Im}\psi=0$ give
\be
\frac{y}{L}=-\tau,
\qquad
\frac{x^2}{L^2}=\tau.
\label{eq:node-example-conditions}
\ee
There are no nodes for $t<0$. At $t=0$ there is a degenerate zero at the origin, while for $t>0$ two nodes appear at
\be
x_\pm=\pm L\sqrt{\tau},
\qquad
y_\pm=-L\tau.
\label{eq:node-pair-trajectories}
\ee
For this solution the Jacobian introduced in \eqref{eq:regular-node-2d} is
\be
J=\frac{2|\psi_0|^2x}{L^3},
\label{eq:node-pair-jacobian}
\ee
so the two nodes carry opposite winding numbers. A contour enclosing both nodes has zero total winding, while the two individual circulation cycles exist only after the critical event and cannot be obtained by continuously transporting corresponding contours from $t<0$ through the creation event.

The density and conserved current remain smooth through the nodal event, so their representation by the gauge-field strengths in \eqref{eq:hydro-gauge-ident-2d} remains well defined even when the density vanishes. By contrast, the reconstruction of the hydrodynamic velocity requires division by the density and therefore becomes singular at the node. Thus, the field-strength representation of the conserved current may remain regular through a topology-changing event, while the full Schrodinger/fluid/gauge correspondence is restricted to regions of nonvanishing density.

The same conclusion applies in $(3+1)$ dimensions. In the present work, we restrict the full correspondence to time intervals between topology-changing nodal events. Extending it through node creation, annihilation or reconnection would require a separate treatment of the vanishing-density locus and the corresponding change in the global phase data.
\endgroup

\section{Topological deformations}
\label{sec:topological-terms}

Topological terms in fluid dynamics have been studied in several formulations, including gauge-theoretic and group-theoretic approaches \cite{Tong:2022gpg,Nastase:2023rou,Eling:2023iyx,Eling:2023apf,Nair:2020kjg,Jackiw:2004nm}. Once the Schrodinger system is written in terms of gauge fields, the same mechanism can be applied. A BF coupling between the hydrodynamic gauge field and an additional one-form shifts the Madelung momentum and thereby produces a nonzero vorticity. We therefore deform the gauge theory side by topological terms and translate the result back into wavefunction variables. Different choices of the additional one-form correspond to different structures in the Schrodinger formulation. An external one-form describes electromagnetic coupling, a dynamical one-form leads to Chern-Simons charge-flux attachment, a composite Berry connection encodes spin and internal polarization, and an intrinsic connection captures holonomy of the phase bundle. In this section, we study these topological deformations and the structures they induce.

\subsection{BF deformation as electromagnetic coupling}

We begin with the simplest deformation, namely the coupling of the gauge-fluid system to an additional abelian one-form $a_\mu$. For simplicity, we work in $(1+2)$ dimensions. In the gauge language, the action (\ref{eq:gauge action for Schr (1+2)}) describes an irrotational quantum fluid. The simplest way to turn on vorticity is to couple $A_\mu$ to $a_\mu$ through a BF term
\be
\label{eq:action for schr + em}
I[A_\mu]
=
\int dt\,d^2x\,
\bigg(
\frac{m\mathbf{E}^2}{2B}
-U(B,\nabla B)
+e\,\varepsilon^{\mu\nu\rho}A_\mu\partial_\nu a_\rho
\bigg).
\ee
The point of this coupling is immediate from the $A_0$ equation of motion. Varying \eqref{eq:action for schr + em} with respect to \(A_0\) gives
\be
\label{eq:mod vort gauge}
\partial_i\!\bigg(\frac{mE^i}{B}\bigg)-eb=0,
\qquad
b\equiv \varepsilon^{ij}\partial_i a_j.
\ee
Using the identification \(\rho=B\) and \(v^i=\varepsilon^{ij}E_j/B\), this becomes
\be
-m\omega-eb=0,
\qquad
\omega=\varepsilon^{ij}\partial_i v_j.
\ee
Thus the BF term sources vorticity by the magnetic field \(b\) of the auxiliary gauge potential \(a_\mu\).

Varying with respect to \(A_i\) yields
\be
\label{eq: gauge EoM Euler}
\partial_t\!\bigg(\frac{mE^i}{B}\bigg)
+\varepsilon^{ki}\partial_k\!\bigg(\frac{m\mathbf E^2}{2B^2}+V+Q\bigg)
+e\,\varepsilon^{ij}e_j=0,
\ee
where
\be
e^i=-\partial_t a^i-\partial_i a^0.
\ee
Contracting \eqref{eq: gauge EoM Euler} with \(\varepsilon^{\ell i}\) and using \eqref{eq:mod vort gauge}, one obtains
\be
\label{eq: Euler with EM}
m\left(\partial_t+v^k\partial_k\right)v^\ell
=
-\partial_\ell(V+Q)
+e\big(b\,\varepsilon^{\ell i}v_i+e^\ell\big).
\ee
This is the Euler equation for a quantum fluid with charge $e$ coupled to an electromagnetic field. The role of the BF term is therefore to introduce the electromagnetic contribution into the fluid dynamics. 

The same mechanism is reflected on the quantum-mechanical side. In the undeformed theory one has \(m\mathbf v=\nabla S\), while \eqref{eq:mod vort gauge} now integrates locally to
\be
\label{eq: v in em}
\mathbf v=\frac1m(\nabla S-e\mathbf a).
\ee
The irrotationality condition is replaced by a twisted potential-flow condition, in which the gauge field shifts the canonical momentum. Once this is recognized, the corresponding Schrodinger equation is
\be
\label{eq:Schr + em}
i\hbar\,\partial_t\psi
=
\bigg[
-\frac{\hbar^2}{2m}\bigg(\nabla-\frac{ie}{\hbar}\mathbf a\bigg)^2
+V+ea^0
\bigg]\psi.
\ee
Applying the Madelung transformation to \eqref{eq:Schr + em} reproduces the continuity equation together with \eqref{eq: Euler with EM}, with velocity given by \eqref{eq: v in em}. Thus the three descriptions remain perfectly aligned. The BF-deformed gauge action, the quantum fluid with nonzero vorticity, and the Schrodinger particle minimally coupled to \(a_\mu\) are three presentations of the same local dynamics.

The global condition is modified accordingly. Since the phase now enters through \(m\mathbf v+e\mathbf a=\nabla S\), single-valuedness of the wavefunction implies
\be
\label{eq:qunt cond with EM}
\oint_C \mathbf v\cdot d\mathbf l
+\frac{e}{m}\oint_C \mathbf a\cdot d\mathbf l
=
\frac{2\pi\hbar n}{m},
\qquad
n\in\mathbb Z.
\ee
Circulation is therefore measured relative to the holonomy of the gauge potential \(a_\mu\).

In \((3+1)\) dimensions the conserved density-current sector is encoded by the two-form gauge field \(B_{\mu\nu}\) and the same BF logic produces the Lorentz force deformation of the Euler equation and the corresponding shifted momentum relation.

\subsection{Chern-Simons term and effective action}

\begingroup

A natural topological deformation in $(2+1)$ dimensions is to add a Chern-Simons term for the hydrodynamic gauge field $A_\mu$. Chern-Simons couplings to nonrelativistic bosonic matter are central to the composite-boson description of the fractional quantum Hall effect \cite{Zhang:1988wy}, to the Chern-Simons Schrodinger system of \cite{Jackiw:1990tz}, and to fluid descriptions of fractional quantum Hall bulk and edge dynamics \cite{Monteiro:2022wip}. Chern-Simons terms also arise in other fluid/gauge systems, including shallow-water theory \cite{Tong:2022gpg} and gauge formulations of incompressible Euler flow \cite{Eling:2023iyx}. Here our purpose is to follow this deformation through the present Schrodinger/fluid/gauge correspondence.
\endgroup

To see what the Chern-Simons deformation does in the gauge formulation of the Schrodinger system, consider the effective gauge action
\be
\label{eq:full chern-simons action}
I[A_\mu]
=
\int dt\,d^2x\,
\bigg(
\frac{m\mathbf{E}^2}{2B}
-U(B,\nabla B)
-\frac{e^2}{2\kappa}\varepsilon^{\mu\nu\rho}A_\mu\partial_\nu A_\rho
\bigg).
\ee
To pass from this gauge description back to Schrodinger variables, one needs the Madelung identification for the momentum. The relevant equation is the Gauss law obtained by varying \eqref{eq:full chern-simons action} with respect to $A_0$
\be
\label{eq: Gauss law CS}
\partial_i\!\bigg(\frac{mE^i}{B}\bigg)
+\frac{e^2}{\kappa}B=0.
\ee
Since $B=\varepsilon^{ij}\partial_iA_j$ and $v^i=\varepsilon^{ij}E_j/B$, one may integrate this relation on any simply connected patch where the wavefunction can be reconstructed and write
\be
\label{eq:CS shifted Madelung via A}
m\mathbf v=\nabla S+\frac{e^2}{\kappa}\mathbf A.
\ee
This is the Chern-Simons-deformed Madelung relation.

In Coulomb gauge \(\partial_iA_i=0\), the gauge field is reconstructed from the density and current (\ref{eq:hydro-gauge-ident-2d}) as
\be
\label{eq: gauge field via psi}
A_i[\psi]
=
-\varepsilon_{ij}\partial_j\Delta^{-1}|\psi|^2,
\qquad
A_0[\psi]
=
\Delta^{-1}\varepsilon_{ij}\partial_i j^j \footnote{Closely related density--flux identifications and Coulomb-gauge
reconstructions of the gauge potential appear in the
Chern-Simons Schrodinger system
\cite{Jackiw:1990tz}.},
\ee
while the current $j^i$ is
\be
\label{eq: current in CS}
j^i=\rho v^i=\frac{|\psi|^2}{m}\left(\partial_iS+\frac{e^2}{\kappa}A_i[\psi]\right).
\ee
Substituting the shifted Madelung relation into the Schrodinger equation gives the nonlocal equation
\be
\label{eq:effective CS nonlocal Schrodinger}
i\hbar\,\partial_t\psi
=
\bigg[
-\frac{\hbar^2}{2m}\bigg(\nabla+\frac{i e^2}{\kappa\hbar}\mathbf A[\psi]\bigg)^2
+V-\frac{e^2}{\kappa}A_0[\psi]
\bigg]\psi,
\ee
so the evolution at a point depends on the density and current over the whole plane through the inverse Laplacian.

Locality can be restored by introducing an auxiliary gauge field \(a_\mu=(a_0,-\mathbf a)\) and considering the action
\be
\label{eq:action for CS EM}
I[A_\mu,a_\mu]
=
\int dt\,d^2x\,
\bigg(
\frac{m\mathbf{E}^2}{2B}
-U(B,\nabla B)
+e\,\varepsilon^{\mu\nu\rho}A_\mu\partial_\nu a_\rho
+\frac{\kappa}{2}\varepsilon^{\mu\nu\rho}a_\mu\partial_\nu a_\rho
\bigg).
\ee
This gauge action is dual to the local Schrodinger equation minimally coupled to a Chern-Simons field
\be
i\hbar\,\partial_t\psi
=
\bigg[
-\frac{\hbar^2}{2m}\bigg(\nabla-\frac{ie}{\hbar}\mathbf a\bigg)^2
+V+ea_0
\bigg]\psi,
\ee
while varying \eqref{eq:action for CS EM} with respect to \(a_\mu\) gives
\be
\label{eq: f and F equal}
\kappa\,\varepsilon^{\mu\nu\rho}f_{\nu\rho}
=
-e\,\varepsilon^{\mu\nu\rho}F_{\nu\rho},
\qquad
f_{\mu\nu}=\partial_\mu a_\nu-\partial_\nu a_\mu.
\ee
Using \eqref{eq:hydro-gauge-ident-2d}, this becomes
\be
\label{eq:cs-current-eq}
\frac{\kappa}{2}\varepsilon^{\mu\nu\rho}f_{\nu\rho}=eJ^\mu,
\qquad
J^\mu=(\rho,\rho v^i)=\left(\psi\psi^*,\frac{\hbar}{2mi}
\bigg[
\psi^*\bigg(\nabla-\frac{ie}{\hbar}\mathbf a\bigg)\psi
-\psi\bigg(\nabla+\frac{ie}{\hbar}\mathbf a\bigg)\psi^*
\bigg]\right).
\ee
In this form the Schrodinger dynamics is local, and the Chern-Simons field \(a_\mu\) is tied directly to the conserved probability current.

One may then pass back to the effective description by integrating out \(a_\mu\). Equation \eqref{eq: f and F equal} fixes only the field strength of \(a_\mu\), so locally it can be solved as
\be
\label{eq:a_mu solution}
a_\mu=-\frac{e}{\kappa}A_\mu+\partial_\mu\lambda,
\ee
with \(\lambda\) a gauge function. Substituting this back into \eqref{eq:action for CS EM} reproduces \eqref{eq:full chern-simons action}, up to a total derivative. The pure \(A_\mu\) theory is therefore the effective description obtained after eliminating the local Chern-Simons gauge field, whereas the two-field formulation keeps that sector explicit and local.

However, this elimination is only local. Equation \eqref{eq:a_mu solution} reconstructs $a_\mu$ only on a simply connected patch, while the two-field theory \eqref{eq:action for CS EM} retains the global phase data through the single-valuedness condition \eqref{eq:qunt cond with EM}. In the effective description the same information reappears through the Gauss law \eqref{eq: Gauss law CS}. To keep the effective theory \eqref{eq:full chern-simons action} in the same topological sector as the underlying local Schrodinger-Chern-Simons system, one must therefore impose the remnant of phase single-valuedness from (\ref{eq:CS shifted Madelung via A})
\be
\label{eq:quant cond in CS}
\oint_C
\left(
 m\mathbf v-\frac{e^2}{\kappa}\mathbf A
\right)\!\cdot d\mathbf l = 2\pi\hbar\,\mathbb Z.
\ee

As an application of the gauge formulation, let us consider the action
\be
\label{eq: nonlinear CS action}
I[A_{\mu}]=\int dt\,d^2x\,\bigg(
\frac{m\mathbf{E}^2}{2B}
-\frac{\hbar^2}{8m}\frac{(\nabla B)^2}{B}-\frac{g}{2}B^2
-\frac{e^2}{2\kappa}\varepsilon^{\mu\nu\rho}A_\mu\partial_\nu A_\rho
\bigg).
\ee
In the two-field formulation this action is dual to the local gauged nonlinear Schrodinger equation coupled to the Chern-Simons gauge field $a_{\mu}$. At the special value $g=\mp \frac{\hbar e^2}{\kappa m}$, the theory supports the soliton sector~\cite{Jackiw:1990tz}. These configurations can be analyzed entirely within the pure gauge theory \eqref{eq: nonlinear CS action}, by deriving the Liouville profile directly from the effective Gauss law and the Bogomolny equations and then evaluating the global condition \eqref{eq:quant cond in CS} on the resulting solution. In this way the gauge formulation reproduces the flux spectrum without the matter field as an independent degree of freedom. A related quantum-hydrodynamic treatment of the Schrodinger-Chern-Simons soliton sector was given in~\cite{Pashaev:2000xq}. The detailed derivation of the gauge soliton solution is deferred to Appendix~\ref{app:selfdual}.

\subsection{Wavefunction representation of Chern-Simons term}

Let us now examine the topological content of the Chern-Simons term from the point of view of the Schrodinger variables. To make this structure explicit, we isolate the Chern-Simons sector of the effective one-field theory and rewrite it directly in terms of the conserved probability current. Using the covariant form of the identification (\ref{eq:hydro-gauge-ident-2d}) 
\be
\label{eq: covariant ident}
J^\mu=-\frac12\varepsilon^{\mu\nu\rho}F_{\nu\rho},
\ee
the Chern-Simons term may be written as
\be
\label{eq: CS term via J}
I_{\rm CS}[A]
=
\frac{e^2}{2\kappa}\int d^3x\,A_\mu J^\mu,
\qquad
A_\mu=(A_0,-\mathbf A).
\ee
The only remaining trace of the gauge formulation in \eqref{eq: CS term via J} is through the gauge field $A_\mu$. To pass to the fluid variables form, one reconstructs $A_\mu$ from $J^\mu$ by inverting \eqref{eq: covariant ident}. After fixing the gauge $\partial_\mu A^\mu=0$, one writes
\be
A_\mu(x)
=
\int d^3y\,K_{\mu\nu}(x-y)\,J^\nu(y),
\ee
where $K_{\mu\nu}$ is the inverse kernel of the curl operator on the transverse subspace,
\be
-\varepsilon^{\mu\alpha\beta}\partial_\alpha^x K_{\beta\nu}(x-y)
=
\left(
\delta^\mu_{\ \nu}
-
\frac{\partial^\mu\partial_\nu}{\Box}
\right)
\delta^{(3)}(x-y).
\ee
Since the current is conserved, $\partial_\mu J^\mu=0$, the transverse projector acts as the identity on $J^\mu$, and the reconstruction gives \eqref{eq: covariant ident}. The remaining freedom in $K_{\mu\nu}$ corresponds to pure-gauge terms and does not affect the functional when contracted with conserved currents, up to boundary terms.

Substituting this reconstruction into the Chern-Simons term (\ref{eq: CS term via J}) gives the bilocal functional
\be
I_{\rm CS}[J]
=
\frac{e^2}{2\kappa}
\int d^3x\,d^3y\,
J^\mu(x)\,K_{\mu\nu}(x-y)\,J^\nu(y).
\label{eq:Hopf-functional}
\ee
This is the Hopf functional associated with the conserved probability current $J^\mu$. In Coulomb gauge this functional takes the following explicit wavefunction form
\begin{multline} \label{eq: CS via psi} I_{\rm CS}[\psi] = \frac{e^2\hbar}{2\pi\kappa m} \int dt\, d^2x\, d^2y\, \operatorname{Im}\!\left( \psi^*(x,t)\partial_i\psi(x,t) \right) |\psi(y,t)|^2 \frac{\varepsilon_{ij}(x-y)_j}{|x-y|^2} \\ - \frac{e^4}{4\pi^2\kappa^2 m} \int dt\, d^2x\, d^2y\, d^2z\, |\psi(x,t)|^2|\psi(y,t)|^2|\psi(z,t)|^2 \\ \times \frac{\varepsilon_{ij}(x-y)_j}{|x-y|^2} \frac{\varepsilon_{ik}(x-z)_k}{|x-z|^2}. 
\end{multline}

To derive this, we fix the Coulomb gauge $\partial_iA^i=0$ and write the Chern-Simons term as
\be
I_{\rm CS}
=
\frac{e^2}{2\kappa}\int dt\,d^2x\,\big(A_0\rho-\mathbf A\cdot\mathbf j\big).
\ee
Using the reconstruction \eqref{eq: gauge field via psi}, the self-adjointness of
$\Delta^{-1}$ and an integration by parts in space, one finds
\be
\int dt\,d^2x\,A_0\rho
=
-\int dt\,d^2x\,\mathbf j\cdot\mathbf A,
\ee
up to a boundary term. Hence
\be
I_{\rm CS}
=
-\frac{e^2}{\kappa}\int dt\,d^2x\,j^i A_i[\psi].
\ee
The current entering this expression is the
Chern-Simons-deformed hydrodynamic current \eqref{eq: current in CS}, namely
\be
j^i
=
j_{\rm can}^i
+
\frac{e^2}{\kappa m}\rho A_i[\psi],
\qquad
j_{\rm can}^i
=
\frac{\hbar}{m}\operatorname{Im}(\psi^*\partial_i\psi).
\ee
Therefore
\be
I_{\rm CS}[\psi]
=
-\frac{e^2}{\kappa}\int dt\,d^2x\,j_{\rm can}^i A_i[\psi]
-
\frac{e^4}{\kappa^2 m}\int dt\,d^2x\,\rho A_i[\psi]A_i[\psi].
\label{eq:planar-Hopf-functional}
\ee
Substituting the explicit Coulomb-gauge kernel in \eqref{eq: gauge field via psi}
and $\rho=|\psi|^2$ gives (\ref{eq: CS via psi}).

The relation between the Chern-Simons term and the Hopf functional is well established in the literature, providing a bridge between topology and observable quantum phases. The Hopf term was shown to generate fractional spin and statistics through the linking of worldlines~\cite{Wilczek:1983cy}, and the Hopf Lagrangian was analyzed in detail in~\cite{Wu:1984kd}. A complementary realization appears in the charge--flux composite description of anyons, where the fractional exchange phase is obtained as the Berry phase accumulated during adiabatic transport~\cite{Arovas:1984qr,Arovas:1985yb}. For a review of fractional spin and statistics in quantum mechanics and field theory, including formulations based on Chern-Simons and Hopf terms, see~\cite{Forte:1990hd}.

Within this established Chern-Simons setting, the point relevant here is the explicit functional obtained after eliminating the hydrodynamic gauge field entirely in favor of the wavefunction variables. In Coulomb gauge this representation also separates naturally into a phase-gradient contribution and a density-dependent dynamical contribution. Indeed, using the integrated Gauss law (\ref{eq:CS shifted Madelung via A}) to reconstruct the gauge field, (\ref{eq:planar-Hopf-functional}) becomes
\be
\label{eq:CS split}
I_{\rm CS}[\psi]=I_{\rm geom}[\psi]+I_{\rm den}[\psi],
\ee
\be
\label{eq: geom and den in Coulomb gauge}
I_{\rm geom}=-\frac{e^2}{\kappa}\int dt\,d^2x\,j_{\rm can}^i A_i[\rho],
\qquad
I_{\rm den}=-\frac{e^4}{\kappa^2m}\int dt\,d^2x\,
\rho A_i[\rho]A_i[\rho].
\ee
The first term contains the phase-gradient part of the wavefunction through $j_{\rm can}^i$ and is linear in the canonical current. It can produce a geometric phase and, in appropriate limits, a topological contribution. The second term contains neither the canonical current nor the phase. Since $A_i[\rho]$ is a nonlocal functional of the density, this term produces an ordinary dynamical phase that depends on the detailed density profile and on the time parametrization of the process. Therefore, within the Coulomb gauge representation, the topological information of the Chern-Simons term is displayed by the geometric part, while the complete wavefunction functional also contains a non-topological density-dependent contribution. We first illustrate this distinction using localized wavepackets and then examine how the decomposition is modified in a more general gauge.

\paragraph{Braiding phase from rigidly transported packets.}

We now illustrate the geometric part of the wavefunction representation of Chern-Simons term on two localized wavepackets on the plane. The centers of the packets follow prescribed slow trajectories $R_1(t)$ and $R_2(t)$. We work in the adiabatic regime, in which the packets remain well separated and approximately rigid. Thus
\be
\rho(x,t)
=
q_1\rho_\sigma(x-R_1(t))
+
q_2\rho_\sigma(x-R_2(t)),
\qquad
\rho_\sigma(x)=\frac{1}{\pi\sigma^2}e^{-|x|^2/\sigma^2}.
\ee
The corresponding transported wavepackets are taken in the form
\be
\psi_a(x,t)
=
\sqrt{q_a\rho_\sigma(x-R_a(t))}
\exp\!\left[
\frac{i}{\hbar}
\Big(m\dot R_a(t)\cdot x+\gamma_a(t)\Big)
\right],
\qquad a=1,2.
\ee
Then
\be
|\psi_a(x,t)|^2
=
q_a\rho_\sigma(x-R_a(t)),
\qquad
j_{{\rm can},a}^i(x,t)
=
q_a\rho_\sigma(x-R_a(t))\,\dot R_a^i(t).
\ee
The braid-dependent first-order contribution comes from \(I_{\rm geom}\). Keeping only the mutual terms and neglecting exponentially small overlaps, one obtains
\begin{multline}
I_{\rm geom}^{(12)}
=
\frac{e^2q_1q_2}{2\pi\kappa}
\int dt\,
\left[
\dot R_1^i\,\mathcal K_i(R_1-R_2)
+
\dot R_2^i\,\mathcal K_i(R_2-R_1)
\right],
\\
\mathcal K_i(R)
=
\int d^2x\,d^2y\,
\rho_\sigma(x)\rho_\sigma(y)
\frac{\varepsilon_{ij}(R+x-y)_j}{|R+x-y|^2}.
\end{multline}
By rotational symmetry the smeared kernel is
\be
\mathcal K_i(R)
=
\left(1-e^{-R^2/(2\sigma^2)}\right)
\frac{\varepsilon_{ij}R_j}{R^2}.
\ee
Since \(\mathcal K_i(-R)=-\mathcal K_i(R)\), this gives
\be
I_{\rm geom}^{(12)}
=
-\frac{e^2q_1q_2}{2\pi\kappa}
\int dt\,
\left(1-e^{-r^2/(2\sigma^2)}\right)\dot\varphi,
\qquad
r_i=(R_1-R_2)_i,
\ee
where
\be
\dot\varphi
=
\frac{\varepsilon_{ij}r_i\dot r_j}{r^2},
\qquad
\varphi=\arctan\frac{r_2}{r_1}.
\ee
Therefore, the geometric term defines the one-form
\be
{\cal A}_{\rm geom}
=
-\frac{e^2q_1q_2}{2\pi\kappa}
\left(1-e^{-r^2/(2\sigma^2)}\right)d\varphi
\ee
on the relative configuration space. For finite \(\sigma\) it is not flat,
\be
d{\cal A}_{\rm geom}
=
-\frac{e^2q_1q_2}{2\pi\kappa}
\frac{r}{\sigma^2}e^{-r^2/(2\sigma^2)}
\,dr\wedge d\varphi,
\ee
so the finite-size phase is geometric but not purely topological. In the large-separation regime \(r\gg\sigma\), however,
\be
I_{\rm geom}^{(12)}
\longrightarrow
-\frac{e^2q_1q_2}{2\pi\kappa}
\int d\varphi.
\ee
The connection then becomes flat away from the coincidence point, and a closed loop depends only on the winding of the relative coordinate. A full braid gives
\be
I_{\rm braid}
=
-\frac{e^2}{\kappa}q_1q_2,
\ee
while a single exchange gives one half of this phase.

The density-dependent part of the Chern-Simons functional gives instead
\be
I_{\rm den}
=
-\int dt\,V_{\rm CS}(R_1-R_2;\sigma),
\qquad
V_{\rm CS}
=
\frac{e^4}{\kappa^2m}
\int d^2x\,\rho(x;R_1,R_2)A_i[\rho](x)A_i[\rho](x).
\ee
This term contains no first-order phase-gradient contribution and no $\dot\varphi$. It depends on the density profile, on the separation, and on the time parametrization of the process. Therefore it contributes a dynamical phase, while the topological contribution is recovered from \(I_{\rm geom}\) in the well-separated adiabatic limit.

\begingroup

\paragraph{Gauge dependence of the geometric--density decomposition.}
The decomposition \eqref{eq:CS split} is defined using the Coulomb-gauge reconstruction \eqref{eq: gauge field via psi}, with a fixed choice of $\Delta^{-1}$ and its boundary conditions. To compare representatives of the same conserved current $J^\mu=(\rho,j^i)$, consider
\be
\label{eq:CS-gauge-transformation}
A_i^{(\zeta)}=A_i^C+\partial_i\chi_\zeta,
\qquad
A_0^{(\zeta)}=A_0^C-\partial_t\chi_\zeta,
\ee
where $\zeta$ labels the gauge choice and $\chi_\zeta$ is smooth and single-valued. We take $\chi_0=0$ for the Coulomb gauge and assume boundary conditions for which the spatial integrations by parts below have no surface contributions. Invariance of the physical velocity in \eqref{eq:CS shifted Madelung via A} requires
\be
\label{eq:CS-gauge-canonical-current}
S^{(\zeta)}=S^C-\frac{e^2}{\kappa}\chi_\zeta,
\qquad
j_{{\rm can},i}^{(\zeta)}
=j_{{\rm can},i}^C
-\frac{e^2}{\kappa m}\rho\,\partial_i\chi_\zeta.
\ee
Thus $\rho$ and $j^i$ are unchanged, whereas the canonical current changes. Using \eqref{eq: CS term via J} and $\partial_t\rho+\partial_i j^i=0$, the change of the full functional is
\be
\label{eq:CS-gauge-boundary}
\begin{aligned}
I_{\rm CS}[A^{(\zeta)}]-I_{\rm CS}[A^C]
&=-\frac{e^2}{2\kappa}\int dt\,d^2x\,
\bigl(\rho\,\partial_t\chi_\zeta
+j^i\partial_i\chi_\zeta\bigr)\\
&=-\frac{e^2}{2\kappa}
\left[\int d^2x\,\rho\,\chi_\zeta\right]_{t_i}^{t_f}.
\end{aligned}
\ee
The functional is therefore invariant when this endpoint term vanishes, for example when $\chi_\zeta$ vanishes at both endpoints.

Now define $I_{\rm geom}^{(\zeta)}$ and $I_{\rm den}^{(\zeta)}$ by using $A_i^{(\zeta)}$ and $j_{{\rm can},i}^{(\zeta)}$ in the two expressions \eqref{eq: geom and den in Coulomb gauge}. Since $j_i=j_{{\rm can},i}^{(\zeta)}+(e^2/\kappa m)\rho A_i^{(\zeta)}$, their sum is $-(e^2/\kappa)\int dt\,d^2x\,j^i A_i^{(\zeta)}$. 
Combining this identity with \eqref{eq:CS-gauge-boundary} gives
\begin{gather}
\label{eq:CS-gauge-decomposition}
I_{\rm CS}[A^{(\zeta)}]
=I_{\rm geom}^{(\zeta)}+I_{\rm den}^{(\zeta)}
+I_{\rm mix}^{(\zeta)}
-\frac{e^2}{2\kappa}
\left[\int d^2x\,\rho\,\chi_\zeta\right]_{t_i}^{t_f},\\
\label{eq:CS-gauge-mix}
I_{\rm mix}^{(\zeta)}
=\frac{e^2}{\kappa}\int dt\,d^2x\,
j^i\partial_i\chi_\zeta
=\frac{e^2}{\kappa}\int dt\,d^2x\,
\chi_\zeta\,\partial_t\rho.
\end{gather}
The last equality again follows from current conservation. These relations show that substituting a general gauge representative into the same two expressions does not, in general, reproduce the complete Chern-Simons functional. The longitudinal contribution and the endpoint term must also be included. For $\chi_0=0$, both vanish and the Coulomb-gauge two-term expression is recovered.

The two pieces are thus separately well defined within the specified Coulomb reconstruction. In particular, $I_{\rm den}[\rho]$ with the Coulomb kernel held fixed remains a functional of the gauge-invariant density; it should be distinguished from $I_{\rm den}^{(\zeta)}$ defined using $A_i^{(\zeta)}$. Their separate geometric and dynamical interpretations refer to the Coulomb-gauge packet calculation above and do not define two independently observable phases. Gauge invariance applies to the closed-cycle phase obtained from the complete functional under the stated boundary and endpoint conditions. In a consistent adiabatic expansion, all contributions at the relevant order in the packet velocities must be retained before identifying the geometric holonomy. Its Coulomb-gauge evaluation gives the braid contribution discussed above, which becomes topological in the well-separated limit.
\endgroup

\subsection{Clebsch variables, Berry connection and spin}

The BF mechanism that introduced an external or dynamical one-form can also be used in a more intrinsic manner. Instead of coupling the hydrodynamic gauge field to an independent \(a_\mu\), one may couple it to a composite one-form built out of scalar fields. The natural hydrodynamic choice is the Clebsch parameterization
\be
\label{eq:Clebsch param}
\tilde A_\mu=\partial_\mu\chi+\beta\,\partial_\mu\alpha.
\ee
The corresponding action of such system is
\be
\label{eq:action for Clebsch}
I[A_\mu,\alpha,\beta]
=
\int dt\,d^2x
\bigg(
\frac{m\mathbf E^2}{2B}
-U(B,\nabla B)
-\varepsilon^{\mu\nu\rho}A_\mu\partial_\nu \tilde A_\rho
\bigg).
\ee
Variation with respect to $A_0$ gives nonzero vorticity
\be
\label{eq: Clebsch vort}
m\omega
=
\tilde B
\equiv
\varepsilon^{ij}\partial_i\tilde A_j
=
-\varepsilon^{ij}\partial_i\beta\,\partial_j\alpha,
\ee
which locally integrates to the familiar Clebsch form of the velocity
\be
m\mathbf v=\nabla S-\beta\,\nabla\alpha,
\ee
after absorbing \(\nabla\chi\) into the phase \(S\).

The virtue of the Clebsch variables is that they package the vorticity sector into a canonical pair. Indeed, variation with respect to \(\alpha\) and \(\beta\) gives
\be
\label{eq:EoM for alpha and beta}
J^\mu\partial_\mu\alpha=0,
\qquad
J^\mu\partial_\mu\beta=0,
\ee
where \(J^\mu=(\rho,\rho\mathbf v)=(B,\varepsilon^{ij}E_j)\). Thus \(\alpha\) and \(\beta\) are comoving scalar labels transported by the mass current. This immediately suggests introducing the second conserved current
\be
\tilde J^\mu
\equiv
-\varepsilon^{\mu\nu\rho}\partial_\nu\beta\,\partial_\rho\alpha.
\ee
Its temporal component is the vorticity density \(\tilde B\), therefore it is the vorticity current. Moreover, because both \(d\alpha\) and \(d\beta\) annihilate \(J^\mu\), and because in \((1+2)\) dimensions the common orthogonal complement to these two one-forms is one-dimensional, the currents \(J^\mu\) and \(\tilde J^\mu\) must be parallel. Matching their time components gives the explicit relation
\be
\tilde J^\mu=\frac{\tilde B}{B}\,J^\mu.
\ee
Equivalently,
\be
\tilde E^i=\frac{\tilde B}{B}E^i=-\varepsilon^{ik}v_k\,\tilde B.
\ee
With this identity the \(A_i\) equation of motion becomes exactly the Euler equation without an extra Lorentz force term. Thus, the Clebsch pair introduces vorticity without deforming the Euler equation.

Clebsch scalars are standard variables in fluid dynamics. In the Schrodinger formulation, however, they are not the most natural internal degrees of freedom. The natural object is a normalized two-component spinor
\be
\label{eq:xi-unit-norm-new}
\xi(x,t)\in\mathbb C^2,
\qquad
\xi^\dagger\xi=1.
\ee
The normalization restricts \(\xi\) to \(S^3\subset\mathbb C^2\), but the overall phase is redundant. Thus the physical target space is
\be
S^3/U(1)\simeq \mathbb{CP}^1\simeq S^2.
\ee
Equivalently, the physical spin texture is encoded by the Bloch vector
\be
\label{eq:n vector}
\mathbf n=\xi^\dagger\boldsymbol{\sigma}\,\xi,
\qquad
\mathbf n^2=1.
\ee

To connect with Clebsch variables, choose a local trivialization of the Hopf fibration \(S^3\to S^2\) and write
\be
\label{eq:xi-local-parameterization}
\xi
=
e^{-i\chi}
\begin{pmatrix}
\sqrt{1-\beta}\\[2mm]
e^{-i\alpha}\sqrt{\beta}
\end{pmatrix},
\qquad
0\leq \beta\leq 1,
\qquad
\alpha\sim \alpha+2\pi,
\qquad
\chi\sim \chi+2\pi.
\ee
Here \(\chi\) is the coordinate along the fiber, while \((\alpha,\beta)\) are local coordinates on the base. On the patch \(\xi_1\neq 0\), the affine coordinate on \(\mathbb{CP}^1\) is
\be
w=\frac{\xi_2}{\xi_1}
=
e^{-i\alpha}\sqrt{\frac{\beta}{1-\beta}},
\label{eq:affine-coordinate-CP1}
\ee
and hence
\be
\beta=\frac{|w|^2}{1+|w|^2},
\qquad
\alpha=-\arg w.
\label{eq:beta-alpha-from-w}
\ee
This shows that the Clebsch pair is nothing but a local coordinate system on $\mathbb{CP}^1$.

It is worth emphasizing the conceptual difference from the usual hydrodynamic use of Clebsch variables. In a generic fluid, \(\alpha\) and \(\beta\) are merely comoving scalar labels, defined only up to area-preserving reparameterizations of the \((\alpha,\beta)\)-plane, and they do not parametrize any preferred target manifold. Here the situation is much more rigid. Once the internal degree of freedom is a normalized spinor, the pair \((\alpha,\beta)\) becomes a local Darboux chart on \(\mathbb{CP}^1\). The corresponding two-form is then not an arbitrary local representative of vorticity, but the pullback of the canonical symplectic form on the coadjoint orbit of \(SU(2)\), namely the Kirillov--Kostant--Souriau form. The fiber phase \(\chi\) is unphysical and only keeps track of the local \(U(1)\) redundancy of the lift from \(S^2\) to \(S^3\).

The natural one-form associated with \(\xi\) is the Berry connection,
\be
\label{eq:Berry connection}
a_\mu\equiv i\hbar\,\xi^\dagger\partial_\mu\xi .
\ee
In our conventions we write $a_\mu=(a_0,-a_i)$, $a^\mu=(a_0,a^i)$,
with \(a^i=\delta^{ij}a_j\). Therefore, the temporal and spatial Euclidean components are 
$a_0=i\hbar\,\xi^\dagger\partial_t\xi$, $a_i=-\,i\hbar\,\xi^\dagger\partial_i\xi$. Since \(\xi^\dagger\xi=1\), differentiation gives \(\partial_\mu(\xi^\dagger\xi)=0\), so \(\xi^\dagger\partial_\mu\xi\) is purely imaginary. It follows that the Berry connection \(a_\mu\) is real.
Substituting \eqref{eq:xi-local-parameterization} gives
\be
a_\mu=\hbar\big(\partial_\mu\chi+\beta\,\partial_\mu\alpha\big),
\ee
so the Clebsch connection is recovered up to the overall factor of \(\hbar\). Under a local phase rotation \(\xi\to e^{i\gamma}\xi\), the one-form transforms as $a_\mu\to a_\mu-\hbar\,\partial_\mu\gamma$,
which in components reads $a_0\to a_0-\hbar\,\partial_t\gamma$,
$a_i\to a_i+\hbar\,\partial_i\gamma$.
Thus, \(a_\mu\) is a genuine \(U(1)\) connection on the Hopf bundle.

With this geometric input the Clebsch theory may be rewritten directly in terms of the spinor
\be
\label{eq:Clebsch action in spin terms}
I[A_\mu,\xi,\Lambda]
=
\int dt\,d^2x
\bigg[
\frac{m\mathbf E^2}{2B}
-U(B,\nabla B)
\bigg]
-\int dt\,d^2x\;
\varepsilon^{\mu\nu\rho}A_\mu\partial_\nu a_\rho
+\int dt\,d^2x\;\Lambda(\xi^\dagger\xi-1),
\ee
where the constraint \(\xi^\dagger\xi=1\) is enforced by the Lagrange multiplier \(\Lambda\). If one stops here, however, the resulting equation for \(\xi\) is only first order in spatial coordinates. To see the physical content of that statement, it is convenient to separate the component along \(\xi\) from the orthogonal one. Introduce projector
\be
P_\perp=1-\xi\xi^\dagger,
\qquad
P_\perp\xi=0,
\qquad
\xi^\dagger P_\perp=0,
\label{eq:Pperp-new}
\ee
so that every \(u\in\mathbb C^2\) decomposes as
\be
u=(\xi^\dagger u)\,\xi+P_\perp u.
\ee
The lift \(\xi\) itself is defined only up to a local \(U(1)\) phase, and therefore \(\partial_\mu\xi\) is not intrinsic to the physical spin space. The intrinsic derivative is the horizontal projection
\be
\label{eq:projected-derivative-new}
\mathfrak D_\mu\xi
\equiv
P_\perp\partial_\mu\xi
=
\partial_\mu\xi+\frac{i}{\hbar}a_\mu\xi .
\ee
With \(a_\mu=(a_0,-a_i)\), this reads in components
\be
\mathfrak D_t\xi=\partial_t\xi+\frac{i}{\hbar}a_0\,\xi,
\qquad
\mathfrak D_i\xi=\partial_i\xi-\frac{i}{\hbar}a_i\,\xi .
\ee
By construction,
\be
\xi^\dagger\mathfrak D_\mu\xi=0.
\ee
Under a local phase rotation \(\xi\to e^{i\gamma}\xi\), together with $a_\mu\to a_\mu-\hbar\,\partial_\mu\gamma$,
the projected derivative transforms covariantly, $\mathfrak D_\mu\xi\to e^{i\gamma}\mathfrak D_\mu\xi$.
Thus, \(\mathfrak D_\mu\) is the \(U(1)\)-covariant derivative.

Varying \eqref{eq:Clebsch action in spin terms} with respect to \(\xi^\dagger\), the component parallel to \(\xi\) is absorbed by \(\Lambda\), while the physical equation is
\be
P_\perp(J^\mu\partial_\mu\xi)
=
J^\mu\mathfrak D_\mu\xi
=
0.
\ee
In the local parameterization \eqref{eq:xi-local-parameterization} this is equivalent to \eqref{eq:EoM for alpha and beta}. Thus \eqref{eq:Clebsch action in spin terms} is a spinorial rewriting of the Clebsch hydrodynamics, but not yet the gauge dual of the spin Schrodinger equation.

Once a two-component wavefunction is written as
\be
\label{eq:spin Madelung}
\Psi=\sqrt{\rho}\,e^{iS/\hbar}\,\xi,
\ee
the Laplacian produces the second order spatial derivative of $\xi$, therefore we need the kinetic term in the gauge action for the spin variable. The natural gauge invariant term is the spin-stiffness \(B\,\mathfrak D_i\xi^\dagger\mathfrak D_i\xi\). The resulting action is
\begin{multline}
\label{eq:minimal spin-gauge action}
I[A_\mu,\xi,\Lambda]
=
\int dt\,d^2x
\Bigg[
\frac{m\mathbf E^2}{2B}
-U(B,\nabla B)
-\frac{\hbar^2}{2m}\,B\,\mathfrak D_i\xi^\dagger\mathfrak D_i\xi
\Bigg]
\\
-\int dt\,d^2x\;
\varepsilon^{\mu\nu\rho}A_\mu\partial_\nu a_\rho
+\int dt\,d^2x\;\Lambda(\xi^\dagger\xi-1).
\end{multline}
This is the gauge theory dual of the free spinor Schrodinger equation
\be
\label{eq:spin-schrodinger}
i\hbar\,\partial_t\Psi
=
\left(
-\frac{\hbar^2}{2m}\nabla^2+V
\right)\Psi,
\ee
with \(\Psi\) given by \eqref{eq:spin Madelung} and Madelung momentum is
\be
m\mathbf v=\nabla S +\mathbf a.
\ee
The detailed derivation is presented in Appendix~\ref{app:derivation}.

\paragraph{Pauli equation.}

This construction extends to the Pauli equation in one higher spatial dimension
\be
i\hbar\,\partial_t\Psi
=
\bigg[
-\frac{\hbar^2}{2m}\bigg(\nabla-\frac{iq}{\hbar}\mathbf A\bigg)^2
+V+q\phi-\mu\,\boldsymbol{\sigma}\!\cdot\!\mathbf B
\bigg]\Psi.
\ee
It admits the gauge-side representation
\begin{multline}
\label{eq: Pauli gauge action}
S[B_{\mu\nu},\xi,\Lambda]
=
\int d^3x\,dt
\bigg[
\frac{mH_iH^i}{2H}
-U(H,\nabla H)
-\frac{\hbar^2}{2m}
H\,\mathfrak D_i\xi^\dagger\mathfrak D^i\xi
\bigg]
+3\int d^3x\,dt\,
\epsilon^{\mu\nu\rho\sigma}B_{\mu\nu}\partial_\rho a_\sigma
\\
-3q\int d^3x\,dt\,
\epsilon^{\mu\nu\rho\sigma}B_{\mu\nu}\partial_\rho A_\sigma
+\mu\int d^3x\,dt\,
H\,\xi^\dagger\boldsymbol{\sigma}\!\cdot\!\mathbf B\,\xi
+\int d^3x\,dt\;\Lambda(\xi^\dagger\xi-1).
\end{multline}
Here \(B_{\mu\nu}\) is the hydrodynamic two-form gauge field of the dual description, while \(A_\mu=(\phi,-\mathbf A)\) is the external electromagnetic potential, with magnetic field \(\mathbf B=\nabla\times\mathbf A\). The equivalence of \eqref{eq: Pauli gauge action} to the Pauli equation follows by repeating the derivation presented in Appendix~\ref{app:derivation}. The quantum hydrodynamic equations of this system are the following
\be
\partial_t(m\mathbf v+q\mathbf A-\mathbf a)
+\nabla\!\left(
\frac{mv^2}{2}
-a_0
+q\phi
+V
+Q
+\frac{\hbar^2}{2m}\mathfrak D_i\xi^\dagger\mathfrak D^i\xi
-\mu\,\xi^\dagger\boldsymbol{\sigma}\!\cdot\!\mathbf B\,\xi
\right)=0,
\ee
where $m\mathbf v+q\mathbf A-\mathbf a=\nabla S$ and the orthogonal spin equation
\be
i\hbar\,\rho\,(\mathfrak D_t+\mathbf v\!\cdot\!\mathfrak D)\xi
=
-\frac{\hbar^2}{2m}\,
P_\perp\!\Big[\partial_i\!\big(\rho\,\mathfrak D_i\xi\big)\Big]
+\frac{i\hbar}{2m}\rho\,\mathbf a\!\cdot\!\mathfrak D\xi
-\mu\,\rho\,P_\perp\!\big[(\boldsymbol{\sigma}\!\cdot\!\mathbf B)\xi\big].
\ee

The hydrodynamical reading of the Pauli equation goes back to Takabayasi \cite{Takabayasi:1954, Takabayasi:1955}, who formulated the dynamics in terms of fluid variables and the gauge invariant spin vector \(\mathbf n\). A complementary group-theoretic formulation of spinning relativistic fluids, based on coadjoint-orbit variables, was developed in \cite{Karabali:2014vla}. The present point of view is slightly different. Here the fundamental variable is the normalized spinor \(\xi\), together with the Berry connection it induces, and the action principle directly yields a covariant equation for \(\xi\). Of course the theory can be rewritten in purely gauge-invariant variables. In particular,
\be
\label{eq:kinetic spin term}
\mathfrak D_i\xi^\dagger\mathfrak D^i\xi
=
\frac14\,\partial_i\mathbf n\cdot\partial^i\mathbf n,
\ee
while the Berry curvature in BF term becomes
\be
f_{\mu\nu}
\equiv
\partial_\mu a_\nu-\partial_\nu a_\mu
=
-\frac{\hbar}{2}\,
\mathbf n\cdot
(\partial_\mu\mathbf n\times\partial_\nu\mathbf n).
\ee
At the level of gauge-invariant observables, one therefore recovers Takabayasi's formulation. Nevertheless, the explicit \((A_\mu,\xi)\) description is more useful for the boundary analysis developed later, because one must distinguish carefully between the ordinary gauge symmetry of the hydrodynamic gauge field and the local \(U(1)\) redundancy of the spinor lift.

In summary, the spinor wavefunction is characterized by variables \((\rho,S,\xi)\). Passing to hydrodynamic variables replaces \(S\) by the velocity, but the spinor \(\xi\) survives unchanged on all sides of the correspondence. On the gauge side it couples to fluid/gauge sector through its Berry connection via the BF term, so the internal spin degree of freedom becomes part of the topological sector of the fluid-gauge correspondence.

\subsection{Non-abelian Berry connection}

Once the abelian Berry connection is recognized as the real object entering the dual action, the next step is to generalize the result to non-abelian connection. In the Born--Oppenheimer point of view the Berry connection lives over the space of slow variables, which in the present context are the spacetime coordinates \(x^\mu=(t,\mathbf x)\) of the hydrodynamic motion. If the slow evolution preserves not a line bundle but a rank-\(k\) degenerate subspace, the relevant connection is non-abelian \cite{Mead:1979ayo, Wilczek:1984dh}. This structure also appears in real-space Berry phases for atoms moving in spatially varying dressed states \cite{Ruseckas:2005vgb, Dalibard:2010ph}. We now adapt this picture to the present setting.

Let \(N\) be the dimension of the ambient internal Hilbert space and \(k\) the rank of the degenerate adiabatic bundle. At each spacetime point choose an orthonormal frame of the \(k\)-plane
\be
Z(x)=\big(\Phi_1(x),\dots,\Phi_k(x)\big)\in\mathbb C^{N\times k},
\qquad
Z^\dagger Z=\mathbf 1_k.
\label{eq:NA-frame}
\ee
The occupied state inside this subspace is described by a reduced wavefunction
\be
\chi(x)\in\mathbb C^k,
\qquad
\Psi_{\rm full}(x)=Z(x)\chi(x).
\label{eq:NA-full-wavefunction-compact}
\ee
The frame is not unique and under a local change of basis
\be
\label{eq:change of fram nonabelian}
Z(x)\to Z(x)U(x),
\qquad
U(x)\in U(k),
\ee
the reduced amplitude transforms as \(\chi\to U^{-1}\chi\). This naturally defines the non-abelian Berry connection
\be
\mathcal A_\mu=i\hbar\,Z^\dagger\partial_\mu Z,
\qquad
\mathcal F_{\mu\nu}
=
\partial_\mu\mathcal A_\nu-\partial_\nu\mathcal A_\mu
-\frac{i}{\hbar}[\mathcal A_\mu,\mathcal A_\nu].
\label{eq:NA-Berry-connection-compact}
\ee
For the spatial components we use the same convention as for the
Abelian gauge fields,
\be
\mathcal A_\mu
=
\bigl(\mathcal A_0,-\mathcal A_i\bigr),
\qquad
\mathcal A_0
=
i\hbar\,Z^\dagger\partial_t Z,
\qquad
\mathcal A_i
=
-\,i\hbar\,Z^\dagger\partial_i Z,
\label{eq:NA-Berry-components}
\ee
where Latin indices $i,j,\ldots$ denote Euclidean spatial components. Under a local change of frame (\ref{eq:change of fram nonabelian}) the non-Abelian Berry connection transforms as
\be
\mathcal A_\mu
\longrightarrow
U^{-1}\mathcal A_\mu U
+
i\hbar\,U^{-1}\partial_\mu U.
\label{eq:NA-Berry-gauge-transformation}
\ee

If the system is also coupled to an external abelian electromagnetic field \(A_\mu^{\text{em}}=(\phi^{\text{em}},-\mathbf A^{\text{em}})\), then the reduced Schrodinger equation is
\be
i\hbar D_t\chi
=
\left[
-\frac{\hbar^2}{2m}D_iD_i
+V\,\mathbf 1_k+\mathcal U
\right]\chi,
\qquad
D_\mu
=
\partial_\mu+\frac{iq}{\hbar}A_\mu^{\text{em}}-\frac{i}{\hbar}\mathcal A_\mu,
\label{eq:NA-Schr-reduced-compact}
\ee
where explicitly spatial indices
\be
D_t
=
\partial_t
+\frac{iq}{\hbar}\phi^{\rm em}
-\frac{i}{\hbar}\mathcal A_0,
\qquad
D_i
=
\partial_i
-\frac{iq}{\hbar}A_i^{\rm em}
+\frac{i}{\hbar}\mathcal A_i,
\label{eq:NA-covariant-derivative-components}
\ee
and \(\mathcal U(x,t)\) is a Hermitian \(k\times k\) matrix potential acting inside the degenerate multiplet.

To reach hydrodynamic variables, we perform the Madelung decomposition of the reduced amplitude
\be
\chi=\sqrt{\rho}\,e^{iS/\hbar}\,\xi,
\qquad
\xi^\dagger\xi=1,
\qquad
\xi\in\mathbb C^k.
\label{eq:NA-Madelung-compact}
\ee
The frame \(Z\) specifies the rank-\(k\) bundle and its non-abelian Berry geometry, while \(\xi\) specifies the occupied direction inside that bundle. The abelian one-form that actually enters the hydrodynamic equations is the projection of the non-abelian connection onto the occupied polarization
\be
a_\mu
\equiv
i\hbar\,\xi^\dagger\partial_\mu\xi
+\xi^\dagger\mathcal A_\mu\xi.
\label{eq:NA-projected-connection-compact}
\ee
This one-form is invariant under \(U(k)\) changes of frame and transforms as an ordinary \(U(1)\) connection under the residual phase rotation of \(\xi\). Introducing also projector
\be
P_\perp=\mathbf 1_k-\xi\xi^\dagger,
\qquad
\mathfrak D_\mu\xi
\equiv
P_\perp\bigg(\partial_\mu-\frac{i}{\hbar}\mathcal A_\mu\bigg)\xi
=
\partial_\mu\xi+\frac{i}{\hbar}(a_\mu-\mathcal A_\mu)\xi,
\label{eq:NA-projected-derivative-compact}
\ee
one finds exactly the same structure as in the abelian spin case. In particular,
\be
m v_i=\partial_i S+a_i-qA_i^{\rm em},
\qquad
\partial_t\rho+\partial_i(\rho v_i)=0.
\label{eq:NA-hydro-identification-compact}
\ee
The Hamilton--Jacobi equation differs from the abelian spin case only through the replacement of the scalar spin interaction by the matrix expectation value \(\xi^\dagger\mathcal U\xi\) and through the appearance of the covariant derivative \(\mathfrak D_i\xi\).

The dual gauge-side construction is therefore immediate. One takes the action of the previous subsection, replaces the abelian Berry one-form by the projected connection \eqref{eq:NA-projected-connection-compact}, and replaces ordinary derivatives acting on \(\xi\) by the projected covariant derivatives \eqref{eq:NA-projected-derivative-compact}. Since the variation is identical in form to the abelian case, we do not repeat it here. The resulting gauge theory is dual to the multicomponent Schrodinger system \eqref{eq:NA-Schr-reduced-compact} on a rank-\(k\) Berry bundle. In the special case \(k=1\), the bundle reduces to a line bundle and one recovers the abelian Berry connection of the previous subsection.

\subsection{The intrinsic holonomy deformation}

There is one further deformation that is conceptually distinct from the previous ones. In the preceding constructions the extra one-form arose either from an external gauge potential or from Berry geometry of an internal degree of freedom. The construction of \cite{Tronci:2020vlq} shows that there is a third possibility. The additional one-form may be an intrinsic \(U(1)\) connection associated with the phase factor of the wavefunction itself. This is useful because it incorporates holonomy directly into the Madelung picture without starting from a multivalued phase.

Following \cite{Tronci:2020vlq}, instead of writing \(\psi=\sqrt\rho\,e^{iS/\hbar}\), one factors the wavefunction as
\be
\psi(x,t)=\sqrt{\rho(x,t)}\,\theta(x,t),
\qquad
\theta(x,t)\in U(1).
\label{eq:holonomy-factor}
\ee
The associated \(U(1)\) connection is
\be
\bar\nu_i\equiv -i\hbar\,\theta^{-1}\partial_i\theta.
\label{eq:holonomy-connection}
\ee
The key step in \cite{Tronci:2020vlq} is to relax the flatness condition only after varying the action. In this way \(\bar\nu_i\) is not forced to remain a pure gradient; its initial curvature may survive as part of the data. One then writes
\be
\bar\nu_i=\partial_i s-\lambda_i,
\qquad
m v_i=\partial_i s-\lambda_i,
\qquad
b_\lambda\equiv \varepsilon^{ij}\partial_i\lambda_j.
\label{eq:holonomy-v}
\ee
Here \(s\) is single-valued, while \(\lambda_i\) carries the intrinsic holonomy.

Note that \(\lambda_i\) is not an external electromagnetic field. It enters the Schrodinger equation through the same coupling, but its meaning is entirely different. It is an internal connection attached to the quantum state and encodes geometric phase data intrinsic to the phase bundle. The resulting equation is
\be
\label{eq:Schr-holonomy-2d}
i\hbar\,\partial_t\psi
=
\left[
\frac{1}{2m}\big(-i\hbar\nabla-\boldsymbol\lambda\big)^2+V+\lambda_0
\right]\psi.
\ee
Since \(s\) is single-valued, the relevant loop relation is now
\be
\oint_C (m\mathbf v+\boldsymbol\lambda)\cdot d\mathbf l
=
\oint_C \nabla s\cdot d\mathbf l
=0.
\label{eq:holonomy-circulation}
\ee
Thus one may have nontrivial smooth curvature without monodromy. If one wishes, one may further admit singular sectors satisfying \(\oint_C \boldsymbol\lambda\cdot d\mathbf l=2\pi\hbar n\), thereby incorporating the topological defects.

This deformation also fits naturally into the gauge description. One simply considers
\be
\label{eq:holonomy-gauge-action}
S[A_\mu,\lambda_\mu]
=
\int dt\,d^2x
\left[
\frac{m\mathbf E^2}{2B}
-U(B,\nabla B)
+\varepsilon^{\mu\nu\rho}A_\mu\partial_\nu\lambda_\rho
\right],
\ee
where \(A_\mu\) is the hydrodynamic gauge field and \(\lambda_\mu=(\lambda_0,-\lambda_i)\) is a convenient spacetime completion of the holonomy one-form. Varying with respect to \(A_0\) gives
\be
\partial_i\left(\frac{mE_i}{B}\right)-b_\lambda=0
\qquad\Longleftrightarrow\qquad
m\omega+b_\lambda=0,
\label{eq:holonomy-gauss}
\ee
which is exactly the gauge form of the statement that the velocity is no longer irrotational. Locally this integrates back to \(m v_i=\partial_i s-\lambda_i\). Variation with respect to \(A_i\) then produces the Euler equation with Lorentz force terms built from the field strength of \(\lambda_\mu\). From the dual point of view, the holonomy deformation is therefore yet another BF deformation.

This gives a useful unifying perspective. In the electromagnetic example the hydrodynamic gauge field couples through a BF term to an external one-form \(a_\mu\). In the Chern-Simons example the one-form is made dynamical and can be integrated out, leaving an induced topological term for \(A_\mu\). In the Clebsch and spin constructions the coupled one-form is composite and is identified with the Berry connection of the spin bundle. In the non-abelian generalization it becomes the projection of a genuinely non-abelian Berry connection onto the occupied polarization. Finally, in the holonomy construction it is an intrinsic connection on the phase bundle of the wavefunction itself. In every case the same general principle is at work: once the density-current sector is encoded by the gauge field \(A_\mu\), any one-form that shifts the Madelung momentum enters the dual theory through a BF coupling to \(A_\mu\).

\section{Edge modes and boundary symmetries}
\label{sec:edge-modes}

The topological terms play the essential role once the system is placed in a space with boundary. In the bulk, BF and Chern-Simons couplings merely deform the equations of motion by shifting the vorticity sector. On a space with boundary, however, they obstruct some gauge transformations from being treated as pure redundancy and thereby convert part of the gauge data into physical edge degrees of freedom. The natural observables are then quasi-local boundary charges, and their Poisson brackets determine the symmetry algebra of the edge theory. We illustrate this mechanism in two representative examples, namely the effective theory with a Chern-Simons term and the spin/Berry realization of the BF coupling.

\subsection{Effective theory with Chern-Simons term}

We begin with the effective action \eqref{eq:full chern-simons action},
\be
I[A]
=
\int_M dt\,d^2x\left[
\frac{m\mathbf E^2}{2B}-U(B,\nabla B)-\frac{\chi}{2}\,\varepsilon^{\mu\nu\rho}A_\mu\partial_\nu A_\rho
\right].
\label{eq:edge-cs-action}
\ee
Without boundary one may discard total derivatives in the variation, but with boundary this is no longer possible. A well-posed variational principle now requires boundary conditions that cancel the boundary contribution while leaving the bulk equations unchanged. Varying the action and integrating by parts gives
\be
\delta I
=
\int_M dt\,d^2x\,{\cal E}^\mu \delta A_\mu
+\int_{\partial M} dt\,ds\, n_\mu S^\mu(\delta A),
\label{eq:edge-cs-var-structure}
\ee
where \({\cal E}^\mu=0\) are the bulk equations of motion and the boundary term is encoded in
\begin{gather}
S^0=-\frac{mE^i}{B}\delta A^i-\frac{\chi}{2}\varepsilon^{ij}\delta A^iA^j, \label{S^0 CS}
\\
S^i=\bigg(-\frac{mE^i}{B}-\frac{\chi}{2}\varepsilon^{ij}A^j\bigg)\delta A_0-\bigg(\frac{mE^2}{2B^2}+V+Q-\frac{\chi}{2}A_0\bigg)\varepsilon^{ij}\delta A^j-\frac{\hbar^2}{4m}\,\frac{\partial_i B}{B}\delta B. \label{S^i CS}
\end{gather}
For a spatial boundary with outward normal \(n_i\), this becomes
\begin{multline}
n_\mu S^\mu(\delta A)=
\left(-m\frac{n_iE^i}{B}-\frac{\chi}{2}\varepsilon^{ij}n_iA_j\right)\delta A_0
-\left(\frac{mE^2}{2B^2}+V+Q-\frac{\chi}{2}A_0\right)\varepsilon^{ij}n_i\delta A_j
\\
-\frac{\hbar^2}{4m}\,\frac{n^i\partial_i B}{B}\,\delta B .
\label{eq:edge-cs-bdry-var}
\end{multline}
A simple choice that removes this boundary term is to fix
\be
A_0\big|_{\partial M}=\text{constant},
\qquad
A_t\big|_{\partial M}=\text{constant},
\label{eq:edge-cs-Dirichlet-A}
\ee
where \(t_i=\varepsilon_{ij}n^j\) is the tangent vector to the boundary and \(A_t=t^iA_i\). The quantum potential contribution allows two options:
\be
B\big|_{\partial M}=\text{constant},
\label{eq:edge-cs-Dirichlet-B}
\ee
or
\be
n^i\partial_iB\big|_{\partial M}=0.
\label{eq:edge-cs-Neumann-B}
\ee
The first condition fixes the boundary density profile, while the second leaves the boundary density free but sets the normal quantum-stress flux to zero. Hence \eqref{eq:edge-cs-Dirichlet-A} together with either \eqref{eq:edge-cs-Dirichlet-B} or \eqref{eq:edge-cs-Neumann-B} defines a good variational problem.

Once a boundary is present, gauge invariance also requires a separate check. Under a gauge transformation \(A_\mu\to A_\mu+\partial_\mu\lambda\) \((\delta A_i=-\partial_i \lambda)\), the Chern-Simons term varies by a total derivative
\be
\delta_\lambda I
=
\int dt\,d^2x\,\partial_\mu N^\mu_\lambda,
\qquad
N^0_\lambda=\frac{\chi}{2}\lambda B,
\qquad
N^i_\lambda=\frac{\chi}{2}\lambda\,\varepsilon^{ij}E_j .
\label{eq:edge-cs-Nmu}
\ee
Therefore
\be
\delta_\lambda I=
-\frac{\chi}{2}\int_{\partial M}dt\,ds\,\lambda\,t^iE_i
=
\frac{\chi}{2}\int_{\partial M}dt\,ds\,\lambda\,\rho v_n .
\label{eq:edge-cs-gauge-var}
\ee
The boundary conditions \eqref{eq:edge-cs-Dirichlet-A} make this variation vanish automatically. Indeed, fixing the boundary values of \(A_0\) and \(A_t\) implies the conductor condition \(E_t|_{\partial M}=0\), and through the gauge-fluid dictionary this is equivalent to the impermeability condition \(v_n|_{\partial M}=0\). Thus, the gauge variation of the action vanishes on the allowed phase space.

\begingroup

The presence of a boundary changes the status of gauge transformations. In the Lagrangian description, the condition \eqref{eq:edge-cs-Dirichlet-A} selects a representative for which the variational principle is well posed. Passing to the canonical description also requires specifying which gauge directions are removed when the constrained phase space is constructed. We retain the boundary gauge orbit of this representative and parametrize it by a scalar $\varphi(t,s)$. Writing the fixed values in \eqref{eq:edge-cs-Dirichlet-A} as $\bar A_0$ and $\bar A_t$, we take
\be
A_0\big|_{\partial M}
=
\bar A_0+\partial_t\varphi,
\qquad
A_t\big|_{\partial M}
=
\bar A_t-\partial_s\varphi .
\label{eq:edge-cs-boundary-orbit}
\ee
The choice $\varphi=0$ reproduces \eqref{eq:edge-cs-Dirichlet-A}. Since $E_t$ and $B$ are gauge invariant, the conditions entering the variational principle remain satisfied throughout the orbit \eqref{eq:edge-cs-boundary-orbit}, and the gauge variation \eqref{eq:edge-cs-gauge-var} continues to vanish.

A gauge transformation acts on the boundary scalar according to
\be
\varphi
\longrightarrow
\varphi+\lambda\big|_{\partial M}.
\label{eq:edge-cs-boundary-phi-transformation}
\ee
After imposing the Gauss constraint (\ref{eq: Gauss law CS}), transformations with $\lambda|_{\partial\Sigma}=0$ remain degenerate gauge directions of the reduced symplectic form and are quotiented out. Transformations with nonzero boundary value act through \eqref{eq:edge-cs-boundary-phi-transformation} and are retained as tangent directions of the boundary phase space. At fixed time they are parametrized by arbitrary functions $\lambda(s)$ on $\partial\Sigma$. These residual gauge directions are the edge modes. Their Hamiltonian generators arise from the boundary contribution to the reduced symplectic form, and the Poisson algebra of these generators gives the edge algebra.

\endgroup

\begingroup

To make edge algebra explicit, we now exhibit the symplectic structure. At fixed time, the canonical symplectic form is
\be
\Omega_\Sigma
=
\int_\Sigma d^2x\,
\delta\Pi^\mu\wedge\delta A_\mu ,
\label{eq:edge-cs-canonical-symplectic}
\ee
where $\Pi^\mu$ is the momentum conjugate to $A_\mu$. After imposing the constraints and quotienting by proper gauge transformations, the resulting symplectic form defines the Poisson bracket on the physical phase space. The residual directions described by \eqref{eq:edge-cs-boundary-orbit} and \eqref{eq:edge-cs-boundary-phi-transformation} are retained in this reduction.

For the action \eqref{eq:edge-cs-action}, the momentum conjugate to $A_i$ follows from the Legendre transformation,
\be
\Pi^i
=
\frac{\partial{\cal L}}
{\partial(\partial_t A_i)}
=
-\frac{mE^i}{B}
-\frac{\chi}{2}\varepsilon^{ij}A_j .
\label{eq:edge-cs-canonical-momentum}
\ee
Since the action contains no time derivative of $A_0$, its conjugate momentum vanishes,
\be
\Pi^0\approx0 .
\label{eq:edge-cs-primary-constraint}
\ee
On this primary constraint surface, the nontrivial part of the equal-time symplectic form is therefore
\be
\Omega_\Sigma^{\rm CS}
=
\int_\Sigma d^2x\,
\delta\Pi^i\wedge\delta A_i .
\label{eq:edge-cs-symplectic-form-def}
\ee
Using (\ref{eq:edge-cs-canonical-momentum}), this becomes
\be
\Omega_\Sigma^{\rm CS}
=
\int_\Sigma d^2x
\left[
-\delta\!\left(\frac{mE^i}{B}\right)\wedge\delta A_i
+
\frac{\chi}{2}\varepsilon^{ij}
\delta A_i\wedge\delta A_j
\right].
\label{eq:edge-cs-symplectic-form}
\ee
The quantum-potential term $U(B,\nabla B)$ contains no time derivatives and therefore does not enter the symplectic form. The term $mE^2/(2B)$ does contribute through $E_i/B$, which is gauge invariant.

The equation obtained previously by varying $A_0$ (\ref{eq: Gauss law CS}) becomes a constraint in the canonical formulation and takes the form
\be
{\cal G}_{\rm CS}
\equiv
\partial_i\Pi^i-\frac{\chi}{2}B
\approx0 .
\label{eq:edge-cs-canonical-gauss}
\ee
Thus the Hamiltonian system is constrained by $\Pi^0\approx0$ and ${\cal G}_{\rm CS}\approx0$.

We now restrict the symplectic form to the Gauss constraint surface. Variations tangent to this surface obey
\be
\delta{\cal G}_{\rm CS}
=
\delta\left(
\partial_i\Pi^i-\frac{\chi}{2}B
\right)
=0 .
\label{eq:edge-cs-linearized-gauss}
\ee
The reduced phase space is obtained by quotienting the proper gauge transformations with
\(\lambda|_{\partial\Sigma}=0\). The residual transformations acting through
\eqref{eq:edge-cs-boundary-phi-transformation} remain tangent directions of the boundary phase space defined by
\eqref{eq:edge-cs-boundary-orbit}. The symplectic form on this reduced phase space is the restriction of (\ref{eq:edge-cs-symplectic-form}) to variations satisfying \eqref{eq:edge-cs-linearized-gauss},
\be
\Omega_{\rm red}^{\rm CS}
=
\left.
\int_\Sigma d^2x
\left[
-\delta\!\left(\frac{mE^i}{B}\right)\wedge\delta A_i
+
\frac{\chi}{2}\varepsilon^{ij}
\delta A_i\wedge\delta A_j
\right]
\right|_{\delta{\cal G}_{\rm CS}=0} .
\label{eq:edge-cs-reduced-symplectic-form}
\ee

We next determine the charge associated with a residual boundary gauge transformation. For a phase-space function $F$, its Hamiltonian vector field $X_F$ is defined by
\be
\Omega_{\rm red}^{\rm CS}(X_F,\delta)
=
-\delta F .
\label{eq:edge-cs-hamiltonian-vector}
\ee
Accordingly, the charge $Q_\lambda^{\rm CS}$ associated with the transformation
\be
\delta_\lambda A_i=-\partial_i\lambda
\label{eq:edge-cs-spatial-gauge-transformation}
\ee
is determined by
\be
\Omega_{\rm red}^{\rm CS}
(\delta_\lambda,\delta)
=
-\delta Q_\lambda^{\rm CS}.
\label{eq:edge-cs-charge-definition}
\ee
For \(\lambda|_{\partial\Sigma}\neq0\), the transformation
\eqref{eq:edge-cs-spatial-gauge-transformation} acts on the boundary coordinate through
\eqref{eq:edge-cs-boundary-phi-transformation}, so that \(\delta_\lambda\) is an admissible tangent direction of the boundary phase space.

Since $E_i/B$ is gauge invariant, variation of canonical momentum gives
\be
\delta_\lambda\Pi^i
=
\frac{\chi}{2}\varepsilon^{ij}\partial_j\lambda .
\label{eq:edge-cs-momentum-gauge-transformation}
\ee
Evaluating the symplectic form on $\delta_\lambda$ and an arbitrary tangent variation $\delta$ gives
\begin{align}
\Omega_\Sigma^{\rm CS}
(\delta_\lambda,\delta)
&=
\int_\Sigma d^2x
\left[
\delta_\lambda\Pi^i\,\delta A_i
-
\delta\Pi^i\,\delta_\lambda A_i
\right]
\nonumber\\
&=
\int_\Sigma d^2x\,
\partial_i\lambda
\left[
\delta\Pi^i
-\frac{\chi}{2}\varepsilon^{ij}\delta A_j
\right].
\label{eq:edge-cs-symplectic-contraction}
\end{align}
Integrating by parts,
\be
\Omega_\Sigma^{\rm CS}
(\delta_\lambda,\delta)
=
\oint_{\partial\Sigma}ds\,
\lambda
\left[
n_i\delta\Pi^i
-\frac{\chi}{2}\varepsilon^{ij}n_i\delta A_j
\right]
-
\int_\Sigma d^2x\,
\lambda\,\delta{\cal G}_{\rm CS}.
\label{eq:edge-cs-symplectic-contraction-boundary}
\ee
The second term vanishes by \eqref{eq:edge-cs-linearized-gauss}. Therefore
\be
\Omega_{\rm red}^{\rm CS}
(\delta_\lambda,\delta)
=
\oint_{\partial\Sigma}ds\,
\lambda
\left[
n_i\delta\Pi^i
-\frac{\chi}{2}\varepsilon^{ij}n_i\delta A_j
\right].
\label{eq:edge-cs-reduced-symplectic}
\ee
For $\lambda|_{\partial\Sigma}=0$ the boundary contribution in
\eqref{eq:edge-cs-reduced-symplectic} vanishes, in agreement with the quotient by proper gauge transformations. For residual transformations acting through
\eqref{eq:edge-cs-boundary-phi-transformation}, the same boundary contribution determines the Hamiltonian charge. Substituting (\ref{eq:edge-cs-canonical-momentum}), we find
\be
\Omega_{\rm red}^{\rm CS}
(\delta_\lambda,\delta)
=
-
\oint_{\partial\Sigma}ds\,
\lambda
\left[
m n_i
\delta\!\left(\frac{E^i}{B}\right)
+
\chi\varepsilon^{ij}n_i\delta A_j
\right].
\label{eq:edge-cs-charge-variation}
\ee
The corresponding charge is
\be
Q_{\lambda}^{\rm CS}
=
\oint_{\partial\Sigma} ds\,
\lambda
\left(
m\frac{E_i}{B}
+
\chi\varepsilon_{ij}A_j
\right)n^i
=
\oint_{\partial\Sigma}
\lambda\left(mv_i-\chi A_i\right)dx^i .
\label{eq:edge-cs-charge}
\ee

Finally, the symplectic form determines the Poisson algebra of these charges. With the convention \eqref{eq:edge-cs-hamiltonian-vector},
\be
\{F,G\}
=
X_G[F]
=
-\Omega_{\rm red}^{\rm CS}(X_F,X_G).
\label{eq:edge-cs-poisson-definition}
\ee
Since $Q_{\lambda}$ generates $\delta_\lambda$, it follows that
\be
\left\{
Q_{\lambda_1}^{\rm CS},
Q_{\lambda_2}^{\rm CS}
\right\}
=
-\Omega_{\rm red}^{\rm CS}
(\delta_{\lambda_1},\delta_{\lambda_2}) .
\label{eq:edge-cs-bracket-symplectic}
\ee
The contribution proportional to $mE^2/(2B)$ vanishes in this pairing because it is gauge invariant. Only the Chern-Simons part contributes,
\begin{align}
\Omega_{\rm red}^{\rm CS}
(\delta_{\lambda_1},\delta_{\lambda_2})
&=
\chi
\int_\Sigma d^2x\,
\varepsilon^{ij}
\partial_i\lambda_1
\partial_j\lambda_2
\nonumber\\
&=
-\chi
\oint_{\partial\Sigma}
\lambda_1\partial_s\lambda_2\,ds ,
\label{eq:edge-cs-algebra-symplectic}
\end{align}
where we used $t_i=\varepsilon_{ij}n^j$. Hence
\be
\left\{
Q_{\lambda_1}^{\rm CS},
Q_{\lambda_2}^{\rm CS}
\right\}
=
\chi
\oint_{\partial\Sigma}
\lambda_1\partial_s\lambda_2\,ds .
\label{eq:edge-cs-algebra}
\ee
Thus, the algebra \eqref{eq:edge-cs-algebra} is the Poisson algebra of the residual boundary gauge directions retained through
\eqref{eq:edge-cs-boundary-orbit} and
\eqref{eq:edge-cs-boundary-phi-transformation}.

\endgroup

The bracket \eqref{eq:edge-cs-algebra} is the affine \(U(1)\), or Kac--Moody, algebra of Chern-Simons theory with boundary. The canonical origin of these edge degrees of freedom was clarified in Chern-Simons-Witten theory \cite{Elitzur:1989nr}, the conformal edge-current interpretation was given in \cite{Balachandran:1991dw}, and the corresponding canonical symmetry-algebra analysis was developed in \cite{Oh:1998sv}. Recent Abelian Chern-Simons edge-mode analyses on manifolds with boundary give the same current-algebra structure \cite{Bertolini:2026qit}. The same algebra appears in the gauge formulation of the \((2+1)\)-dimensional incompressible Euler equation \cite{Eling:2023iyx}, and closely related edge structures arise in the gauge-theoretic description of shallow water \cite{Tong:2022gpg,Sheikh-Jabbari:2023eba}.

An affine \(U(1)\) algebra strongly suggests a chiral boundary description. In many familiar examples the same current algebra is realized by a chiral boson and is accompanied by conformal symmetry and a Virasoro structure. The topological part of the present theory indeed behaves in that way. If one isolates the pure Chern-Simons sector and chooses temporal gauge \(A_0=0\), its bulk constraint makes the spatial connection locally flat, so near the boundary one may write \(A_i=\partial_i\phi\). Substituting into the topological action gives
\be
I^{\rm CS}_{\rm edge}
=
\frac{\chi}{2}\int_{\partial\Sigma}dt\,ds\,\partial_s\phi\,\partial_t\phi,
\label{eq:edge-cs-toplag}
\ee
which is the action of a two-dimensional chiral boson.

\begingroup

In the Schrodinger problem, however, the current algebra is only the equal-time kinematics. It does not by itself imply that the complete boundary dynamics is conformal, because the Hamiltonian still contains the non-topological Schrodinger terms. In particular, the quantum potential introduces higher spatial derivatives and therefore produces dispersive dynamics rather than the linear chiral dispersion characteristic of a conformal edge mode.
\endgroup

This is already visible in the bulk linearized theory. For this illustrative calculation, we first set the Chern-Simons coupling to zero and linearize the undeformed theory. Expanding around a static configuration with constant density \(\rho=\rho_0+\delta\rho\) and vanishing velocity, one finds
\be
Q=
-\frac{\hbar^2}{4m\rho_0}\nabla^2\delta\rho+O(\delta\rho^2),
\label{eq:edge-cs-Q-linear}
\ee
so the linearized continuity and Euler equations are
\be
\partial_t\delta\rho+\rho_0\,\partial_i\delta v^i=0,
\qquad
m\,\partial_t\delta v_i
=
\partial_i\frac{\hbar^2}{4m\rho_0}\nabla^2\delta\rho.
\label{eq:edge-cs-linear-eqs}
\ee
Eliminating \(\delta v_i\) gives
\be
\partial_t^2\delta\rho
+\frac{\hbar^2}{4m^2}\nabla^4\delta\rho
=0,
\label{eq:edge-cs-dispersive-wave}
\ee
and therefore the dispersion relation
\be
\omega^2=\frac{\hbar^2}{4m^2}k^4 .
\label{eq:edge-cs-bogoliubov}
\ee

\begingroup

When \(\chi\neq0\), the background must instead satisfy the Chern-Simons Gauss law. Taking the curl of the shifted Madelung relation \eqref{eq:CS shifted Madelung via A} gives
\be
m\,\bar\omega=\chi\,\bar\rho,
\qquad
\bar\omega\equiv\varepsilon^{ij}\partial_i\bar v_j,
\label{eq:edge-cs-background-gauss}
\ee
so a homogeneous configuration with \(\bar\rho\neq0\) and \(\bar{\mathbf v}=0\) is not an admissible background. A constant-density background must carry nonzero vorticity. 

The Chern-Simons deformation makes the emergent Lorentz-force contribution cancel against the vorticity term, leaving the usual quantum Euler equation together with the additional vorticity constraint. Linearizing about a stationary background \((\bar\rho,\bar v_i)\) satisfying \eqref{eq:edge-cs-background-gauss} gives
\begin{gather}
\partial_t\delta\rho
+\partial_i\!\left(\bar\rho\,\delta v^i+\delta\rho\,\bar v^i\right)=0,
\label{eq:edge-cs-linear-cont-background}
\\
m\left[(\partial_t+\bar v^j\partial_j)\delta v_i
+\delta v^j\partial_j\bar v_i\right]
=-\partial_i\delta Q,
\label{eq:edge-cs-linear-euler-background}
\\
m\,\delta\omega=\chi\,\delta\rho,
\qquad
\delta\omega\equiv\varepsilon^{ij}\partial_i\delta v_j.
\label{eq:edge-cs-linear-gauss-background}
\end{gather}
For a locally constant background density \(\bar\rho=\rho_0\), the quantum-potential fluctuation is still (\ref{eq:edge-cs-Q-linear}).

Thus, the Chern-Simons coupling changes the kinematics by tying the transverse velocity fluctuation to the density fluctuation through \eqref{eq:edge-cs-linear-gauss-background}, and background vorticity introduces the corresponding advection/shear terms in \eqref{eq:edge-cs-linear-euler-background}. It does not remove the two additional spatial derivatives coming from the quantum potential. The precise normal-mode spectrum depends on the chosen Chern-Simons-compatible stationary background and boundary conditions, but the conclusion relevant for the edge theory is unchanged. The affine algebra is fixed by the topological term, whereas the Hamiltonian contains higher-derivative dispersive terms. Hence, the presence of the Kac--Moody algebra does not imply that the full edge theory is a conformal chiral theory.
\endgroup

A complete boundary reduction, together with a direct interpretation in terms of boundary density and phase fluctuations of the Schrodinger field, will not be attempted here.

\subsection{Boundary analysis of the spin BF theory}
\label{subsec:edge-spin}

We now turn to the system with spin degree of freedom. Compared with the Chern-Simons example, the structure is slightly richer because the topological term is now of BF type and involves two one-forms: the hydrodynamic gauge field \(A_\mu\) and the Berry connection \(a_\mu\). Correspondingly, one must distinguish two different local symmetries: the ordinary gauge transformation of \(A_\mu\) and the fiber \(U(1)\) redundancy of the Hopf lift \(\xi\to e^{i\gamma/\hbar}\xi\).

We start from the minimal action
\begin{multline}
I[A_\mu,\xi,\Lambda]
=
\int_M dt\,d^2x\Bigg[
\frac{m\mathbf E^2}{2B}
-U(B,\nabla B)
-\frac{\hbar^2}{2m}\,B\,\mathfrak D_i\xi^\dagger\mathfrak D_i\xi
\Bigg]
-
\\
\int_M dt\,d^2x\;\varepsilon^{\mu\nu\rho}A_\mu\partial_\nu a_\rho
+\int_M dt\,d^2x\;\Lambda(\xi^\dagger\xi-1).
\label{eq:edge-spin-action}
\end{multline}
Repeating the same variational analysis as before gives
\be
\label{eq:S^0 spin}
S^0=-\frac{mE^i}{B}\delta A^i+\varepsilon^{ij}A_i\delta a_j
\ee
and
\begin{multline}
\label{eq:S^i spin}
S^i=-\frac{mE^i}{B}\delta A_0-\bigg(\frac{mE^2}{2B^2}+V+Q+\frac{\hbar^2}{2m}\mathfrak D_k\xi^\dagger\mathfrak D_k\xi\bigg)\varepsilon^{ij}\delta A^j-\frac{\hbar^2}{4m}\,\frac{\partial_i B}{B}\delta B+
\\
\varepsilon^{ij}(A_0\delta a_j-A_j\delta a_0)-\frac{\hbar^2}{2m}B\bigg[(\mathfrak D_i\xi)^\dagger\delta\xi+\delta\xi^\dagger\mathfrak D_i\xi-\xi^{\dagger}\delta\xi\cdot(\mathfrak{D}_i\xi)^{\dagger}\xi+\delta\xi^{\dagger}\xi\cdot \xi^{\dagger}\mathfrak{D}_i\xi\bigg].
\end{multline}
In the gauge field sector we impose the same conditions as earlier \eqref{eq:edge-cs-Dirichlet-A} together with either \eqref{eq:edge-cs-Dirichlet-B} or \eqref{eq:edge-cs-Neumann-B}. The new ingredient is the spin sector. Inspecting \eqref{eq:S^i spin}, one sees that the additional boundary conditions required for a well-posed variational principle are
\begin{gather}
 a_0\big|_{\partial M}=\text{constant},
\qquad
 a_t\big|_{\partial M}=\text{constant},
\label{eq:bc for a}
\\
n^i\mathfrak D_i\xi\big|_{\partial M}=0.
\label{eq:edge-spin-Neumann-xi}
\end{gather}
The first condition fixes the variation of the Berry connection, while the second is a covariant Neumann condition for the spinor. The variations appearing in \eqref{eq:bc for a} are of course not independent of \(\delta\xi\), since \(a_\mu\) is itself composite.

\begingroup

We now turn to the equal-time symplectic structure and algebra of gauge charges. The canonical symplectic form on a spatial slice is
\be
\Omega_\Sigma^{\rm spin}
=
\int_\Sigma d^2x
\left[
\delta\Pi_A^i\wedge\delta A_i
+
\delta\pi_\xi\wedge\delta\xi
+
\delta\pi_{\xi^\dagger}\wedge\delta\xi^\dagger
\right],
\label{eq:edge-spin-canonical-symplectic}
\ee
where contraction over the spinor components is understood. The momentum conjugate to the hydrodynamic gauge field is
\be
\Pi_A^i
=
-\frac{mE^i}{B}.
\label{eq:edge-spin-PiA}
\ee

The theory is constrained. The equation of motion obtained by varying $A_0$ gives the Gauss constraint
\be
{\cal G}_A
\equiv
\partial_i\Pi_A^i
-
\varepsilon^{ij}\partial_i a_j
\approx0.
\label{eq:edge-spin-canonical-gauss}
\ee
The spin variables are restricted by the normalization $\xi^\dagger\xi=1$. Accordingly, variations tangent to the constraint surface satisfy
\be
\delta{\cal G}_A=0,
\qquad
\delta(\xi^\dagger\xi)=0.
\label{eq:edge-spin-constraint-variations}
\ee

We first consider the hydrodynamic gauge symmetry,
\be
\delta_\lambda A_0=\partial_t\lambda,
\qquad
\delta_\lambda A_i=-\partial_i\lambda.
\label{eq:edge-spin-A-transformation}
\ee
The boundary treatment of the hydrodynamic gauge field is the same as in the Chern-Simons case. The Dirichlet condition \eqref{eq:edge-cs-Dirichlet-A} selects the representative $\varphi=0$ of the boundary orbit \eqref{eq:edge-cs-boundary-orbit}, while transformations with nonzero $\lambda|_{\partial\Sigma}$ act through \eqref{eq:edge-cs-boundary-phi-transformation} and are retained as tangent directions of the boundary phase space. Transformations with $\lambda|_{\partial\Sigma}=0$ remain proper gauge redundancies and are quotiented out. 

Since the transformation \eqref{eq:edge-spin-A-transformation} acts only on $A_\mu$ and leaves $\xi$ unchanged, its charge may be evaluated using a local representative of the Hopf fiber. For this calculation we choose the horizontal representative
\be
\xi^\dagger\delta\xi=0.
\label{eq:edge-spin-horizontal-variation}
\ee
This condition fixes the local fiber variation while leaving the projective spin degrees of freedom unrestricted. Substituting $a_i=-i\hbar\,\xi^\dagger\partial_i\xi$ into the spin contribution $\varepsilon^{ij}A_i\delta a_j$ in \eqref{eq:S^0 spin} and integrating spatially by parts, the additional terms are proportional to $\xi^\dagger\delta\xi$ and vanish on \eqref{eq:edge-spin-horizontal-variation}. On this representative the spin momenta are
\be
\pi_\xi
=
i\hbar\varepsilon^{ij}A_i\partial_j\xi^\dagger,
\qquad
\pi_{\xi^\dagger}
=
-i\hbar\varepsilon^{ij}A_i\partial_j\xi.
\label{eq:edge-spin-Pixi}
\ee
Together with \eqref{eq:edge-spin-PiA}, this gives
\be
\Omega_\Sigma^{\rm spin}
=
\int_\Sigma d^2x
\Bigg[
-\delta\!\left(\frac{mE^i}{B}\right)\wedge\delta A_i
+
i\hbar\varepsilon^{ij}
\delta\!\left(A_i\partial_j\xi^\dagger\right)\wedge\delta\xi
-
i\hbar\varepsilon^{ij}
\delta\!\left(A_i\partial_j\xi\right)\wedge\delta\xi^\dagger
\Bigg].
\label{eq:edge-spin-symplectic-form}
\ee
It is useful to introduce
\be
\Xi_j(\delta)
=
i\hbar\left[
(\partial_j\xi^\dagger)\delta\xi
-
\delta\xi^\dagger\partial_j\xi
\right],
\label{eq:edge-spin-Xi}
\ee
for which
\be
\varepsilon^{ij}\partial_i\Xi_j
=
\delta\left(
\varepsilon^{ij}\partial_i a_j
\right).
\label{eq:edge-spin-Xi-curvature}
\ee
Contracting \eqref{eq:edge-spin-symplectic-form} with $\delta_\lambda$ and integrating by parts gives
\be
\Omega_\Sigma^{\rm spin}(\delta_\lambda,\delta)
=
\oint_{\partial\Sigma}ds\,
\lambda
\left[
n_i\delta\Pi_A^i
-
n_i\varepsilon^{ij}\Xi_j(\delta)
\right]
-
\int_\Sigma d^2x\,
\lambda\,\delta{\cal G}_A.
\label{eq:edge-spin-A-contraction}
\ee
The bulk term vanishes on the constraint surface. From the definition of the Berry connection
\be
\Xi_j(\delta)
=
\delta a_j
+
i\hbar\partial_j(\xi^\dagger\delta\xi).
\label{eq:edge-spin-Xi-a}
\ee
On the horizontal representative \eqref{eq:edge-spin-horizontal-variation}, \eqref{eq:edge-spin-Xi-a} reduces to $\Xi_j=\delta a_j$. The boundary condition \eqref{eq:bc for a} then removes the spin contribution, leaving
\be
\Omega_{\rm red}^{\rm spin}(\delta_\lambda,\delta)
=
-\oint_{\partial\Sigma}ds\,
\lambda m n_i
\delta\!\left(\frac{E^i}{B}\right).
\label{eq:edge-spin-A-reduced}
\ee
With the same convention for Hamiltonian generators as in the Chern-Simons analysis
\be
\Omega_{\rm red}^{\rm spin}(\delta_\lambda,\delta)
=
-\delta Q^A_\lambda,
\ee
we obtain surface charge
\be
Q^A_\lambda
=
m\oint_{\partial\Sigma}ds\,
\lambda\frac{E_i}{B}n^i
=
m\oint_{\partial\Sigma}\lambda v_i\,dx^i.
\label{eq:edge-spin-QA}
\ee
Since $E_i/B$ is invariant under the $A_\mu$ gauge transformation,
\be
\{Q^A_{\lambda_1},Q^A_{\lambda_2}\}=0.
\label{eq:edge-spin-QA-algebra}
\ee

We next consider the local $U(1)$ redundancy of the Hopf lift,
\be
\delta_\gamma\xi=\frac{i}{\hbar}\gamma\,\xi,
\qquad
\delta_\gamma\xi^\dagger=-\frac{i\gamma}{\hbar}\xi^\dagger
\qquad
\delta_\gamma a_0=-\partial_t\gamma,
\qquad
\delta_\gamma a_i=+\partial_i\gamma,
\qquad
\delta_\gamma A_\mu=0.
\label{eq:edge-spin-fiber-transformation}
\ee
The fixed condition \eqref{eq:bc for a} selects one representative of the boundary fiber orbit. A transformation with boundary-dependent $\gamma$ need not preserve this representative, so in examining the fiber direction we retain the corresponding orbit and write
\be
a_0\big|_{\partial M}
=
\bar a_0-\partial_t\vartheta,
\qquad
a_t\big|_{\partial M}
=
\bar a_t+\partial_s\vartheta .
\label{eq:edge-spin-boundary-fiber-orbit}
\ee
Under \eqref{eq:edge-spin-fiber-transformation},
\be
\vartheta
\longrightarrow
\vartheta+\gamma\big|_{\partial M}.
\label{eq:edge-spin-boundary-fiber-transformation}
\ee
The curvature $da$ and the covariant condition \eqref{eq:edge-spin-Neumann-xi} are unchanged along this orbit.

To determine whether the fiber transformation carries an independent charge, we return to the spin variables before imposing the horizontal choice \eqref{eq:edge-spin-horizontal-variation}, since $\delta_\gamma\xi$ is itself vertical. Using the unreduced presymplectic form together with (\ref{eq:edge-spin-fiber-transformation}), its contraction with an arbitrary tangent variation reduces to terms proportional to $\delta(\xi^\dagger\xi)$, $\partial_i(\xi^\dagger\xi)$, and $\partial_i\delta(\xi^\dagger\xi)$. These vanish identically on the normalized-spinor phase space. Hence, the fiber transformation is a null direction of the unreduced presymplectic form,
\be
\Omega_{\rm pre}^{\rm spin}(\delta_\gamma,\delta)=0.
\label{eq:edge-spin-fiber-null}
\ee
Accordingly, the fiber $U(1)$ does not possess an independent surface charge
\be
Q_\gamma=0.
\label{eq:edge-spin-no-fiber-charge}
\ee
The boundary fiber orbit \eqref{eq:edge-spin-boundary-fiber-orbit} is therefore quotiented out, and \eqref{eq:edge-spin-horizontal-variation} provides a convenient local representative of the resulting spin phase space.

In an ordinary BF theory the bulk term accompanying the total derivative is related to the second Gauss constraint associated with the independent BF gauge field. Here there is no such constraint, because $a_\mu$ is composite. This distinction between the two BF formulations is discussed in \cite{Eling:2023apf}. When both fields entering the BF term are independent gauge fields, the equal-time symplectic form contains the mixed canonical pairing
\be
\Omega_\Sigma^{\rm BF}
\sim
\int_\Sigma
\varepsilon^{ij}\delta A_i\wedge\delta a_j,
\ee
and the two temporal components impose two independent Gauss constraints. Both gauge transformations then possess surface charges, and their mixed bracket gives the BF boundary algebra
\cite{Balachandran:1992qg,Balachandran:1993wj,Amoretti:2012hs}. This is not the canonical structure of the present theory. When the second one-form is replaced by composite variables, the second gauge-field constraint and the corresponding charge are absent. The spin theory belongs to the latter class as far as its boundary charge structure is concerned.

The same result can be checked geometrically using the local parametrization introduced above. The Berry connection takes the form
\be
a=\hbar\left(d\chi+\beta\,d\alpha\right),
\qquad
da=\hbar\,d\beta\wedge d\alpha.
\label{eq:edge-spin-local-check}
\ee
The fiber coordinate $\chi$ therefore drops out of the BF coupling, confirming directly that its local shift does not provide an additional canonical direction. The two-form $d\beta\wedge d\alpha$ is the local expression for the symplectic form on $\mathbb{CP}^1\simeq S^2$. A theory depending only on this two-form has the familiar Clebsch symplectomorphism invariance discussed in \cite{Eling:2023apf}. In the present spin theory, however, the stiffness term \eqref{eq:kinetic spin term} also depends on the Fubini--Study metric on $\mathbb{CP}^1$, so generic symplectomorphisms are not symmetries. The fiber $U(1)$ remains exact, but \eqref{eq:edge-spin-fiber-null} shows that it remains a gauge redundancy rather than generating a second boundary charge.

Thus, the spin theory has a single family of boundary charges, \eqref{eq:edge-spin-QA}, with the abelian algebra \eqref{eq:edge-spin-QA-algebra}. The mixed boundary algebra characteristic of BF theory with two independent gauge fields is absent.

\endgroup

A complete study of the resulting edge dynamics, including the effect of the stiffness and quantum potential terms on the boundary Hamiltonian and excitation spectrum, will be left for future work. Although the analysis was carried out in $(2+1)$ dimensions, it extends straightforwardly to higher dimensions with the same conclusion.

\section{Sound, acoustic memory and the infrared triangle}
\label{sec:memory}

The gauge description developed above becomes especially useful once the Schrodinger system supports a genuine acoustic sector. Then one may ask under what conditions an outgoing acoustic disturbance leaves behind a permanent late-time imprint, in close analogy with the acoustic memory effects discussed for ordinary sound waves in \cite{deAguiarAlves:2025vfu}. However, for the linear Schrodinger equation the infrared sector is too dispersive for the memory logic to apply.

We work on the hydrodynamic side and linearize the continuity \eqref{eq: continuity} and Euler equations \eqref{eq: Euler eq} around a static homogeneous background. In the purely linear case, this reduces to the linearized system already displayed in \eqref{eq:edge-cs-linear-eqs}. It is convenient to encode the irrotational velocity perturbation by the phase potential
\be
\varphi\equiv \frac{\delta S}{m},
\qquad
\delta\mathbf v=\nabla\varphi.
\label{eq:phase-potential-def}
\ee
The corresponding displacement memory is then the time-integrated velocity kick,
\be
\Delta X_i(\mathbf x)
\equiv
\int_{-\infty}^{+\infty}dt\,\delta v_i(t,\mathbf x)
=
\partial_i\int_{-\infty}^{+\infty}dt\,\varphi(t,\mathbf x).
\label{eq:displacement-memory-def}
\ee
Any infrared memory effect must therefore come from a low-frequency, radiative contribution to \(\varphi\). At this point the linear Schrodinger theory immediately runs into an obstruction. As we already saw in the discussion of the edge dynamics, if one keeps only the quantum potential contribution then the linearized excitations obey a quartic dispersive equation, and the corresponding low-momentum behavior is \(\omega\sim k^2\) \eqref{eq:edge-cs-bogoliubov}. The sector is therefore not acoustic, i.e. there is no branch of the form \(\omega\sim c_s k\), no analogue of a null far zone for propagating sound, and hence no ordinary infrared memory effect of the familiar kind.

The situation changes once the Schrodinger equation is deformed by a local nonlinear interaction. In hydrodynamic variables that deformation appears as a barotropic pressure term and leads to the Bogoliubov dispersion relation
\be
\omega^2
=
c_s^2k^2+\frac{\hbar^2}{4m^2}k^4,
\qquad
c_s^2=\frac{\rho_0 g}{m},
\label{eq:bogoliubov-disp}
\ee
with \(g\) the nonlinear coupling. The long-wavelength part of \eqref{eq:bogoliubov-disp} is now linear, so the theory contains a genuine sound mode.

\paragraph{Retarded solution and the memory regime}

We now consider the nonlinear Schrodinger system in hydrodynamic variables
\be
\partial_t\rho+\nabla\cdot(\rho\mathbf v)=0,
\qquad
m(\partial_t+\mathbf v\cdot\nabla)\mathbf v=-\nabla(g\rho+Q+V),
\label{eq:3d-madelung-memory}
\ee
and linearize around a homogeneous static background
\be
\rho=\rho_0+\delta\rho,
\qquad
\mathbf v=\delta\mathbf v.
\label{eq:3d-linearization-memory}
\ee
The linearized equations are
\be
\partial_t\delta\rho+\rho_0\nabla\cdot\delta\mathbf v=0,
\label{eq:3d-linearized-continuity-memory}
\ee
\be
m\partial_t\delta v_i
=
-g\partial_i\delta\rho
+\frac{\hbar^2}{4m\rho_0}\partial_i\nabla^2\delta\rho-\partial_i V.
\label{eq:3d-linearized-euler-memory}
\ee
Eliminating \(\delta\rho\) in favor of the phase potential \(\varphi\) defined in \eqref{eq:phase-potential-def}, we obtain
\be
\left(
\partial_t^2-c_s^2\nabla^2+\frac{\hbar^2}{4m^2}\nabla^4
\right)\varphi
=
-\frac{1}{m}\partial_t V.
\label{eq:3d-phi-wave-memory}
\ee
Thus the memory problem is governed by the retarded Green function of the linearized wave operator, which contains both the acoustic piece and the higher-derivative quantum correction,
\be
\varphi(t,\mathbf x)
=
\int dt'\,d^3x'\,
G_R(t-t',\mathbf x-\mathbf x')J_V(t', \mathbf x'),
\label{eq:retarded-solution-new}
\ee
where $J_V(t,\mathbf x)=-\frac{1}{m}\partial_t V(t, \mathbf x)$. The relevant regime is the simultaneous infrared and far-zone limit. Introducing the healing length
\be
\xi\equiv \frac{\hbar}{2mc_s},
\label{eq:healing-length-memory}
\ee
we impose
\be
k\xi\ll 1,
\qquad
r\gg \xi.
\label{eq:memory-regime}
\ee
The infrared condition \(k\xi\ll1\) ensures that the Bogoliubov dispersion relation \eqref{eq:bogoliubov-disp} is dominated by its linear acoustic part, while the far-zone condition \(r\gg\xi\) suppresses the short-range contribution associated with the healing-length scale.

The retarded Green function in this case can be found explicitly. 
The Fourier transform of the retarded kernel is
\be
\widetilde G_R(\omega,\mathbf k)
=
\frac{1}{c_s^2k^2+\alpha k^4-(\omega+i0)^2},
\label{eq:GR-omega-k}
\ee
where $\alpha\equiv \frac{\hbar^2}{4m^2}$. The position-space kernel \(\widetilde G_R(\omega;r)\), with \(r=|\mathbf x-\mathbf x'|\), is then obtained by integrating over momentum
\be
\widetilde G_R(\omega;r)
=
\int \frac{d^3k}{(2\pi)^3}\,
\frac{e^{i\mathbf k\cdot(\mathbf x-\mathbf x')}}{c_s^2k^2+\alpha k^4-(\omega+i0)^2}.
\label{eq:GR-omega-r-int}
\ee
After carrying out the angular integral and factorizing the denominator as
\be
c_s^2k^2+\alpha k^4-(\omega+i0)^2
=
\alpha\,(k^2-k_\omega^2)(k^2+\kappa_\omega^2),
\ee
one can evaluate the remaining radial integral by partial fractions, which yields
\be
\widetilde G_R(\omega;r)
=
\frac{1}{4\pi r\sqrt{c_s^4+4\alpha(\omega+i0)^2}}
\left(e^{ik_\omega r}-e^{-\kappa_\omega r}\right),
\label{eq:green-exact}
\ee
where
\be
k_\omega^2
=
\frac{-c_s^2+\sqrt{c_s^4+4\alpha(\omega+i0)^2}}{2\alpha},
\qquad
\kappa_\omega^2
=
\frac{c_s^2+\sqrt{c_s^4+4\alpha(\omega+i0)^2}}{2\alpha}.
\label{eq:green-kappa-k}
\ee
The square roots are chosen on the retarded branches $\operatorname{Im}k_\omega\geq0$,
$\operatorname{Re}\kappa_\omega>0$. The first term in \eqref{eq:green-exact} is the propagating acoustic contribution, whereas the second is an evanescent piece controlled by the healing length and exponentially suppressed when \(r\gg\xi\). At small frequency one has \(\kappa_\omega\simeq \xi^{-1}\), so the Yukawa term is exponentially suppressed as \(e^{-r/\xi}\). In the regime \eqref{eq:memory-regime}, the Green function therefore reduces to the retarded acoustic propagator
\be
\widetilde G_R(\omega;r)
\simeq
\frac{e^{i\omega r/c_s}}{4\pi c_s^2r}.
\label{eq:green-memory-regime}
\ee
If the source spectrum is concentrated at frequencies
\be
|\omega|\lesssim \omega_*,
\qquad
\omega_*\ll \frac{c_s}{\xi},
\label{eq:source-soft-scale}
\ee
or, equivalently, if its characteristic duration obeys
\be
T_{\rm src}\sim \omega_*^{-1}\gg \frac{\xi}{c_s},
\label{eq:source-soft-time}
\ee
then only the acoustic part of the propagator contributes appreciably. For the memory observable using \eqref{eq:displacement-memory-def} together with
\eqref{eq:retarded-solution-new}, one obtains
\bea
\Delta X_i(\mathbf x)
&=&
\partial_i\int_{-\infty}^{+\infty}dt\,\varphi(t,\mathbf x)
\notag\\
&=&
\partial_i\int d^3x'\int\frac{d\omega}{2\pi}
\left(\int_{-\infty}^{+\infty}dt\,e^{-i\omega t}\right)
\widetilde G_R(\omega;\mathbf x-\mathbf x')\,
\widetilde J_V(\omega,\mathbf x')
\notag\\
&=&
\partial_i\int d^3x'\,
\widetilde G_R(0;\mathbf x-\mathbf x')\,
\widetilde J_V(0,\mathbf x'),
\label{eq:memory-zero-mode}
\eea
where, with the Fourier convention used above
\be
\widetilde J_V(\omega,\mathbf x)
=
\frac{i\omega}{m}\widetilde V(\omega,\mathbf x),
\qquad
J_V(t,\mathbf x)
=
-\frac{1}{m}\partial_t V(t,\mathbf x).
\label{eq:J-source}
\ee
The memory functional therefore projects the response directly onto the zero-frequency
sector. If the external potential has well-defined early- and late-time limits
\be
V(t,\mathbf x)\longrightarrow V_\pm(\mathbf x),
\qquad
t\to\pm\infty,
\ee
then
\be
\widetilde J_V(0,\mathbf x)
=
\int_{-\infty}^{+\infty}dt\,J_V(t,\mathbf x)
=
\frac{1}{m}\left[V_-(\mathbf x)-V_+(\mathbf x)\right].
\label{eq:JV-zero-mode}
\ee
Thus, a potential which returns to the same asymptotic value gives no zero-frequency
source component in this channel, while a quench-like potential can produce a static
zero-mode response.

However, there is a distinction between this static zero-mode response and the
leading radiative memory. In the acoustic far zone the contribution of
\(J_V\) to the leading field is
\be
\varphi_V(t, \mathbf r)
=
\frac{1}{4\pi c_s^2 r}\,
\mathcal J_V(u,\hat{\mathbf r})
+
O\!\left(\frac{1}{r^2},\frac{e^{-r/\xi}}{r}\right),
\qquad
u=t-\frac{r}{c_s},
\ee
where
\be
\mathcal J_V(u,\hat{\mathbf r})
=
\int d^3x'\,
J_V\!\left(u+\frac{\hat{\mathbf r}\cdot\mathbf x'}{c_s},\mathbf x'\right).
\ee
This may equivalently be written as
\be
\varphi_V^{(1)}(u,\hat{\mathbf r})
=
-\frac{1}{4\pi m c_s^2}
\partial_u\mathcal V(u,\hat{\mathbf r}),
\qquad
\mathcal V(u,\hat{\mathbf r})
=
\int d^3x'\,
V\!\left(u+\frac{\hat{\mathbf r}\cdot\mathbf x'}{c_s},\mathbf x'\right).
\label{eq:phiV-one}
\ee
Therefore, a potential with finite early- and late-time limits gives
\be
\Delta\varphi_V^{(1)}(\hat{\mathbf r})=0.
\ee
It may still produce a nonzero displacement through the zero-frequency formula
\eqref{eq:memory-zero-mode}, but that response is Coulombic in the far zone.

To describe the radiative memory sector, we now introduce an effective acoustic source \(J_{\rm eff}\),
\be
\left(
\partial_t^2-c_s^2\nabla^2+\frac{\hbar^2}{4m^2}\nabla^4
\right)\varphi
=
J_{\rm eff}.
\label{eq:Jeff-wave}
\ee
Here \(J_{\rm eff}\) denotes the effective source seen after projecting the near-zone dynamics onto the soft acoustic branch. It includes the potential contribution $-\frac{1}{m}\partial_t V$ and may also contain contributions associated with boundary driving, incoming acoustic data, image sources implementing boundary conditions, or nonlinear near-zone dynamics. We do not assume that such additional contributions are generated generically by the nonlinear Schrodinger dynamics. The analysis below is conditional on the asymptotic behavior of \(J_{\rm eff}\), and determines the memory whenever its retarded moment has a nonzero early--late jump.

In the far zone,
\be
\label{eq: far zone asympt}
\varphi(t,r,\hat{\mathbf r})
=
\frac{\varphi^{(1)}(u,\hat{\mathbf r})}{r}
+
O\!\left(\frac{1}{r^2},\frac{e^{-r/\xi}}{r}\right),
\ee
with
\be
\varphi^{(1)}(u,\hat{\mathbf r})
=
\frac{1}{4\pi c_s^2}
\mathcal J_{\rm eff}(u,\hat{\mathbf r}),
\qquad
\mathcal J_{\rm eff}(u,\hat{\mathbf r})
=
\int d^3x'\,
J_{\rm eff}\!\left(u+\frac{\hat{\mathbf r}\cdot\mathbf x'}{c_s},\mathbf x'\right).
\label{eq:Jeff-retarded-moment}
\ee
The leading radiative memory is therefore
\be
\Delta\varphi^{(1)}(\hat{\mathbf r})
=
\frac{1}{4\pi c_s^2}
\Delta\mathcal J_{\rm eff}(\hat{\mathbf r}).
\label{eq:phi-one-memory}
\ee
Accordingly, the phase acquires the asymptotic shift
\be
\Delta S(t,r,\hat{\mathbf r})
=
\frac{m}{r}\Delta\varphi^{(1)}(\hat{\mathbf r})
+
O\!\left(\frac{1}{r^2},\frac{e^{-r/\xi}}{r}\right),
\label{eq:phase-memory}
\ee
and the leading displacement memory is
\bea
\Delta X_i^{\rm rad}
&=&
\int_{-\infty}^{+\infty}dt\,\partial_i\varphi(t,r,\hat{\mathbf r})
\notag\\
&=&
-\frac{\hat r_i}{c_s r}
\Delta\varphi^{(1)}(\hat{\mathbf r})
+
O\!\left(\frac{1}{r^2},\frac{e^{-r/\xi}}{r}\right).
\label{eq:displacement-memory}
\eea
Thus, the radiative memory is controlled by the jump of the leading asymptotic
phase coefficient. Equivalently,
\be
\Delta\varphi^{(1)}(\hat{\mathbf r})
=
-c_s r\,\hat r_i\Delta X_i^{\rm rad}
+
O(r^{-1}).
\label{eq:phi-displacement-relation}
\ee
A leading $1/r$ acoustic memory therefore requires
\be
\Delta\mathcal J_{\rm eff}(\hat{\mathbf r})\neq0.
\label{eq:Jeff-memory-condition}
\ee

The retarded Green-function analysis gives the standard acoustic response of the Bogoliubov mode and is closely related to the sound, Bose--Einstein-condensate and superfluid memory analyses \cite{deAguiarAlves:2025vfu,Datta:2020rrf,Jensen:2026hrx}. We now translate this hydrodynamic memory into the dual gauge description. In $(3+1)$ dimensions the relevant variables are those of the two-form field, whose asymptotic data encode the same late-time phase shift and displacement memory. This representation further shows that the zero-average part of the memory is generated by a residual large gauge transformation, whereas the monopole component remains an independent global mode.

\paragraph{The two-form gauge description of acoustic memory}

The \((1+3)\)-dimensional hydrodynamic formulation is written in terms of a two-form gauge potential \(B_{\mu\nu}\), with field strength
\be
H_{\mu\nu\rho}=3\partial_{[\mu}B_{\nu\rho]},
\label{eq:exact-two-form-memory}
\ee
defined modulo \(B_{\mu\nu}\to B_{\mu\nu}+2\partial_{[\mu}\varepsilon_{\nu]}\) and \(\varepsilon_\mu\to\varepsilon_\mu+\partial_\mu\lambda\). The hydrodynamic variables are identified by
\be
\rho=\varepsilon^{ijk}H_{ijk},
\qquad
\rho v^i=-3\varepsilon^{ijk}H_{0jk},
\label{eq:exact-hydro-ident-memory}
\ee
so the Bianchi identity is exactly the continuity equation.

To extract the acoustic sector, we linearize around the homogeneous static background \(\rho=\rho_0\), \(v^i=0\), or equivalently \(H_{\mu\nu\rho}=\bar H_{\mu\nu\rho}+h_{\mu\nu\rho}\) with \(\bar H_{ijk}=\frac{\rho_0}{6}\varepsilon_{ijk}\) and \(\bar H_{0ij}=0\). Then
\be
\delta\rho=\varepsilon^{ijk}h_{ijk},
\qquad
\rho_0\,\delta v^i=-3\varepsilon^{ijk}h_{0jk},
\label{eq:h-linearized-ident-memory}
\ee
so the perturbations entering the acoustic memory formula are exactly the linearized field strengths of the dual two-form theory. In the radiation region one may therefore write locally
\be
h_{\mu\nu\rho}=3\partial_{[\mu}b_{\nu\rho]},
\label{eq:h-from-b}
\ee
and analyze the far zone in retarded coordinates \(u=t-r/c_s\), \(x^i=r\hat r^{\,i}(x^A)\).
The acoustic branch then propagates with the effective metric
\be
ds^2
=
-c_s^2dt^2+dr^2+r^2\gamma_{AB}dx^Adx^B
=
-c_s^2du^2-2c_s\,du\,dr+r^2\gamma_{AB}dx^Adx^B,
\label{eq:metric-retarded}
\ee
where \(\gamma_{AB}\) is the metric on the unit sphere and \(D_A\) the associated covariant derivative. The full nonlinear Schrodinger theory does not possess an exact relativistic null infinity, but in the acoustic regime \eqref{eq:memory-regime} the outgoing signal is still organized along approximately constant \(u\), so this large-\(r\) description is the natural one for the memory problem.

Using the far-zone asymptotics (\ref{eq: far zone asympt}) and identifications (\ref{eq:phase-potential-def}), (\ref{eq:h-linearized-ident-memory}),
we may now translate the phase memory into gauge variables. One finds that the leading radiative component of the field strength in coordinates (\ref{eq:metric-retarded}) is
\be
h_{uAB}
=
r\,N_{AB}(u,\hat{\mathbf r})
+O\!\left(1,e^{-r/\xi}\right),
\qquad
N_{AB}=\frac{\rho_0}{6c_s}\,\varepsilon_{AB}\,\partial_u\varphi^{(1)}.
\label{eq:radiative-two-form}
\ee
In a gauge where the radiative data are carried by the angular two-form
\be
b_{AB}=r\,b^{(0)}_{AB}(u,\hat{\mathbf r})+O(1),
\label{eq:bAB-expansion}
\ee
so that
\be
N_{AB}=\partial_u b^{(0)}_{AB}.
\label{eq:N-b-relation}
\ee
Integrating over retarded time gives the permanent change of the leading angular field, i.e. memory tensor
\be
\Delta_{AB}
\equiv
\Delta b^{(0)}_{AB}
=
\frac{\rho_0}{6c_s}\,\varepsilon_{AB}\,\Delta\varphi^{(1)}(\hat{\mathbf r}).
\label{eq:delta-b0}
\ee
This is the gauge-theory representative of the leading radiative memory effect. Indeed, using \eqref{eq:phi-displacement-relation} gives
\be
\Delta_{AB}
=
-\frac{\rho_0}{6}\,r\,\varepsilon_{AB}\,\hat r^{\,i}\Delta X_i^{\rm rad}
+O(r^{-1}).
\label{eq:delta-b-displacement}
\ee
Therefore the leading phase shift, the radiative displacement memory and the two-form memory tensor are the three ways of packaging the same observable infrared data.

\begingroup

We now compare this tensor with residual large gauge transformations. Since the leading angular two-form scales as in \eqref{eq:bAB-expansion}, a transformation that acts on the leading radiative datum must have
\be
\varepsilon_A(u,r,x^B)
=
r\,\varepsilon_A^{(0)}(x^B)+O(1).
\label{eq:large-gauge-parameter-falloff}
\ee
The $O(r)$ scaling refers to covariant coordinate components and corresponds to finite angular components in an orthonormal frame. It then follows that
\be
\delta_\varepsilon b^{(0)}_{AB}
=
2D_{[A}\varepsilon^{(0)}_{B]}.
\label{eq:large-gauge-on-B}
\ee
Thus an $O(1)$ coordinate component $\varepsilon_A$ would affect only the subleading part of $b_{AB}$ and would not shift the leading memory datum.
\endgroup

\begingroup

On $S^2$, a smooth one-form may be decomposed as $\varepsilon_A^{(0)}=D_A\lambda+\varepsilon_A{}^B D_B\chi$. The gradient part does not contribute to \eqref{eq:large-gauge-on-B}, since $D_{[A}D_{B]}\lambda=0$. This is precisely the asymptotic manifestation of the gauge-for-gauge redundancy $\varepsilon_\mu\to\varepsilon_\mu+\partial_\mu\lambda$ of the two-form potential. Since $H^1(S^2)=0$ \footnote{Here $H^p(S^2)$ denotes the real de Rham cohomology of the sphere, i.e. closed $p$-forms modulo exact ones. For $S^2$, $H^1(S^2)=0$ and $H^2(S^2)\simeq\mathbb{R}$. Thus there are no nontrivial harmonic one-forms, while the sphere volume form represents the nontrivial two-form cohomology class.}, there is no harmonic one-form contribution, and the nontrivial large gauge transformations may therefore be represented by the coexact part $\varepsilon_A{}^B D_B\chi$.
\endgroup

To make contact with this form, decompose the sphere function \(\Delta\varphi^{(1)}\) into its average and zero-average pieces
\be
\Delta\varphi^{(1)}(\hat{\mathbf r})
=
D^2\chi(\hat{\mathbf r})+M,
\qquad
M\equiv \frac{1}{4\pi}\int_{S^2}d\Omega\,\Delta\varphi^{(1)}.
\label{eq:phase-decomposition}
\ee

\begingroup

Using the identity\footnote{Let
$v_A=\varepsilon_A{}^C D_C\chi$. Since
$D_A\varepsilon_{BC}=0$ and $D_AD_B\chi=D_BD_A\chi$, the antisymmetric
tensor $2D_{[A}v_{B]}$ is proportional to $\varepsilon_{AB}$. Contracting
with $\varepsilon^{AB}/2$ gives
$\varepsilon^{AB}D_Av_B
=\varepsilon^{AB}\varepsilon_B{}^C D_AD_C\chi
=-D^2\chi$, which proves \eqref{eq:sphere-identity-gauge}.}
\endgroup

\be
2D_{[A}\big(\varepsilon_{B]}{}^C D_C\chi\big)
=
-\varepsilon_{AB}D^2\chi,
\label{eq:sphere-identity-gauge}
\ee
we can rewrite the memory tensor as
\be
\Delta_{AB}
=
2D_{[A}\varepsilon^{(0)}_{B]}
+\frac{\rho_0}{6c_s}\,M\,\varepsilon_{AB},
\qquad
\varepsilon^{(0)}_A
\equiv
-\frac{\rho_0}{6c_s}\,\varepsilon_A{}^B D_B\chi.
\label{eq:late-early-B}
\ee
The first term is exactly the variation generated by a large gauge transformation. Therefore, the zero-average part of the memory tensor is itself a large gauge shift
\be
\Delta^{\rm exact}_{AB}=\delta_\varepsilon b^{(0)}_{AB}.
\label{eq:memory-is-gauge}
\ee
If \(b_{AB}^{(0)-}\) and \(b_{AB}^{(0)+}\) denote the early- and late-time asymptotic configurations, then
\be
b^{(0)+}_{AB}-b^{(0)-}_{AB}
=
\delta_\varepsilon b^{(0)}_{AB}
+\frac{\rho_0}{6c_s}\,M\,\varepsilon_{AB}.
\label{eq:vacuum-transition}
\ee

\begingroup

Thus, radiation drives a transition between asymptotic configurations. The zero-average part of this transition is generated by a large gauge transformation, while the monopole term has a different status. The two-form $M\varepsilon_{AB}$ is proportional to the harmonic volume form on $S^2$ and represents the nontrivial class in $H^2(S^2)$. In particular, the integral of $2D_{[A}\varepsilon^{(0)}_{B]}$ over the sphere vanishes for a globally defined gauge parameter, whereas the integral of $M\varepsilon_{AB}$ is nonzero for $M\neq0$. The monopole mode therefore cannot be generated by a residual large gauge transformation and represents an independent global zero mode of the memory.
\endgroup

The memory/asymptotic-symmetry part of the acoustic infrared triangle therefore persists in nonlinear Schrodinger theory in the following sense. The full theory admits an exact two-form reformulation, but its asymptotic memory interpretation emerges only after one linearizes around the homogeneous vacuum and projects to the soft acoustic branch. In that regime, the field $\varphi^{(1)}$ controls the radiative displacement memory \eqref{eq:displacement-memory} and the radiative two-form data \eqref{eq:radiative-two-form}. Its zero-average part also parametrizes the large-gauge transition \eqref{eq:memory-is-gauge}, while the monopole component represents additional global data. Thus, the zero-average part of the memory and the asymptotic symmetry are organized by the same infrared mode, as in the sound analysis \cite{deAguiarAlves:2025vfu}.

A brief comment on the soft-theorem corner is in order. In the dual two-form language, the charges associated with these large gauge transformations are expected to obey Ward identities whose momentum-space form is the scalar soft theorem, in close analogy with the discussions of scalar and two-form asymptotic symmetry \cite{Campiglia:2017dpg, Campiglia:2018see, Francia:2018jtb}. We will not attempt that derivation here. For our purposes, the point is simply that in the linearized acoustic sector that gives rise to the memory effect, the asymptotic data associated with the large gauge transformation are also the data entering the soft limit.

\section{Conclusion}
\begingroup

What emerges from this analysis is a separation between the kinematic information fixed by current conservation, the global information carried by the wavefunction phase, and the genuinely dynamical content of the Schrodinger system. The gauge variables make this separation particularly transparent. On regions of nonvanishing density they reorganize the hydrodynamic equations without changing their local physical content, while the reconstruction of a wavefunction still requires the global winding data. The same viewpoint also exposes the limitation of this description: when the density vanishes, the hydrodynamic variables become singular, and evolution through changes of nodal topology lies outside the sector treated here.

A common pattern also emerges among the different topological couplings. External gauge fields, Berry data, spin variables and phase-bundle holonomy all modify the relation between mechanical momentum and the phase of the wavefunction, and the BF structure provides a uniform way of implementing these modifications. The Chern-Simons sector illustrates a different aspect of the correspondence. Eliminating the hydrodynamic gauge potential produces a nonlocal wavefunction functional whose Coulomb-gauge form separates transport-sensitive and density-sensitive contributions. This separation is useful for isolating the adiabatic braiding phase, but the two terms are not independent gauge-invariant observables. Away from the Coulomb representative the longitudinal and endpoint contributions must be restored, while finite wavepackets retain dynamical corrections that disappear only in the appropriate well-separated limit.

The boundary problem brings out a further distinction between topological kinematics and Hamiltonian dynamics. In the Chern-Simons sector the boundary current algebra is fixed by the equal-time symplectic structure, whereas the non-topological terms control the actual edge evolution and retain the dispersive scale associated with the quantum potential. The spin/BF system shows that the field content is equally important. Because the Berry one-form is composite, its fiber redundancy does not supply the additional canonical boundary generator that would be present in a BF theory built from two independent gauge fields. Thus, the boundary symmetry structure depends not only on the form of the topological coupling but also on which fields represent independent directions in phase space.

The nonlinear acoustic regime provides a different use of the gauge description. Once the Bogoliubov branch becomes sound-like at long wavelengths, its late-time response can be encoded in the asymptotic data of the dual two-form. The angular component with vanishing sphere average lies along a residual gauge orbit, whereas the monopole component represents separate global information. This establishes the link between acoustic memory and asymptotic gauge symmetry in the sector considered here. A derivation of the corresponding charge Ward identity and soft theorem would be needed to complete the infrared triangle.

Several extensions are suggested by these results. It would be useful to understand how the gauge description can be continued through nucleation, annihilation or reconnection of nodes, to derive the boundary Hamiltonian and spectrum of the surviving edge degrees of freedom, and to formulate the soft Ward identity directly in the two-form variables. Extensions to many-body, dissipative and spinful systems should further clarify which parts of this structure survive once the simple single-particle Madelung description is no longer available.
\endgroup

\appendix

\section{Conventions for Lorentz and spatial notation}
\label{app:notation}

Throughout, spacetime points are written as
\be
x^\mu=(t,x^i).
\ee
In \((2+1)\) dimensions we use
\be
\eta_{\mu\nu}=\mathrm{diag}(1,-1,-1),
\qquad
i,j=1,2,
\ee
while in \((3+1)\) dimensions we use
\be
\eta_{\mu\nu}=\mathrm{diag}(1,-1,-1,-1),
\qquad
i,j,k=1,2,3.
\ee
Once we pass to purely spatial indices, they are always raised and lowered with the Euclidean metric \(\delta_{ij}\). Thus, for any spatial vector,
\be
v_i=\delta_{ij}v^j.
\ee
In particular, for the electromagnetic potential we write
\be
A^\mu=(A_0,\mathbf A),
\qquad
A_\mu=(A_0,-\mathbf A),
\ee
where the components of the spatial vector \(\mathbf A\) satisfy \(A_i=A^i\) because the spatial indices are Euclidean. Likewise,
\be
\partial_\mu=(\partial_t,\partial_i)=(\partial_t,\nabla),
\qquad
\partial^\mu=(\partial_t,-\partial_i)=(\partial_t,-\nabla).
\ee
With these conventions, a gauge transformation acts as
\be
A_\mu\to A_\mu+\partial_\mu\lambda,
\ee
so its spatial components transform as
\be
A_i\to A_i-\partial_i\lambda,
\ee
not with a plus sign. The same bookkeeping applies to the Berry connection,
\be
a_\mu=i\hbar\,\xi^\dagger\partial_\mu\xi.
\ee
Accordingly,
\be
a^\mu=(a_0,a_i),
\qquad
a_\mu=(a_0,-a_i),
\ee
and, using the derivative conventions above, the spatial component is
\be
a_i=-\,i\hbar\,\xi^\dagger\partial_i\xi.
\ee

Our Levi-Civita conventions are
\be
\varepsilon^{012}=+1,
\qquad
\varepsilon^{12}=+1,
\qquad
\varepsilon^{0123}=+1,
\qquad
\varepsilon^{123}=+1.
\ee
The contraction identities used repeatedly in the main text are
\be
\varepsilon^{ij}\varepsilon_{kj}=\delta^i{}_k
\qquad \text{in \((2+1)\) dimensions},
\ee
and
\be
\varepsilon^{ijk}\varepsilon_{\ell jk}=2\,\delta^i{}_{\ell}
\qquad \text{in \((3+1)\) dimensions}.
\ee

As a simple illustration of these conventions, in \((2+1)\) dimensions let
\be
F_{\mu\nu}=\partial_\mu A_\nu-\partial_\nu A_\mu,
\qquad
J^\mu=-\frac12\,\varepsilon^{\mu\nu\rho}F_{\nu\rho}=-\varepsilon^{\mu \nu \rho}\partial_{\nu}A_{\rho}.
\ee
Then
\be
J^0=-\varepsilon^{0ij}\partial_i(-A_j)
=\varepsilon^{ij}\partial_iA_j
\equiv B,
\ee
while for the spatial current
\be
J^i=-(\varepsilon^{i0j}\partial_t(-A_j)+\varepsilon^{ij0}\partial_j A_0)
=-(-\varepsilon^{ij}\partial_t(-A_j)+\varepsilon^{ij}\partial_j A_0)=\varepsilon^{ij}(-\partial_t A_j-\partial_j A_0)\equiv \varepsilon^{ij}E_j.
\ee
This is the identification used in the main text. Whenever both Lorentz and spatial formulas appear in the same computation, these are the conventions assumed.

\section{Derivation of soliton solution}
\label{app:selfdual}

This appendix derives the soliton solution of theory
\be
\label{eq: nonlinear CS action app}
I[A_{\mu}]=\int dt\,d^2x\,\bigg(
\frac{m\mathbf{E}^2}{2B}
-\frac{\hbar^2}{8m}\frac{(\nabla B)^2}{B}-\frac{g}{2}B^2
-\frac{e^2}{2\kappa}\varepsilon^{\mu\nu\rho}A_\mu\partial_\nu A_\rho
\bigg).
\ee
and shows how the global condition \eqref{eq:quant cond in CS} reproduces the flux quantization.

To identify soliton solutions, we look for static field configurations of finite total energy. The stress-energy tensor on the gauge side is obtained by coupling the theory to a background metric and setting
\be
T_{\mu\nu}
=
-\frac{2}{\sqrt{-g}}
\frac{\delta(\sqrt{-g}\mathcal L)}{\delta g^{\mu\nu}}
=
-2\frac{\delta\mathcal L}{\delta g^{\mu\nu}}
+g_{\mu\nu}\mathcal L.
\ee
Since the Chern-Simons term is metric-independent, it does not contribute to the energy. A quick check using
\begin{gather}
E^2
=
-F_{0i}F^{0i}
=
-\Big(F_{0i}F_{0j}g^{00}g^{ij}
+F_{0j}F_{ik}g^{0i}g^{jk}\Big),
\\[1mm]
2B=g^{ij}g^{kl}\varepsilon_{ik}F_{jl},
\end{gather}
shows that, among the remaining terms, only the electric contribution depends on \(g^{00}\). Thus the energy of a static configuration is
\be
\label{eq:energy}
E
=
\int d^2x
\bigg(
\frac{m\mathbf E^2}{2B}
+\frac{\hbar^2}{8m}\frac{(\nabla B)^2}{B}
+\frac{g}{2}B^2
\bigg).
\ee

The Gauss law of the effective theory reads
\be
\label{eq: Gauss law CS app}
\partial_i\!\bigg(\frac{mE^i}{B}\bigg)
+\frac{e^2}{\kappa}B=0.
\ee
Using \eqref{eq: Gauss law CS app}, one may complete the square in the Bogomolny fashion:
\be
\frac{m\mathbf{E}^2}{2B}
+\frac{\hbar^2}{8m}\frac{(\nabla B)^2}{B}
=
\frac{m}{2B}
\bigg(E^i\mp\frac{\hbar}{2m}\partial_i B\bigg)^2
\pm \frac{e^2\hbar}{2\kappa m}B^2
\pm \partial_i\bigg(\frac{\hbar E^i}{2}\bigg).
\ee
For well-behaved fields the total derivative does not contribute to the integrated energy. Exactly as in \cite{Jackiw:1990tz}, at the special value
\be
g=\mp \frac{\hbar e^2}{\kappa m}
\ee
the energy becomes a sum of squares and is minimized by the first-order self-duality equation
\be
\label{eq: self-dual eq}
E_i=\sigma\, \frac{\hbar}{2m}\,\partial_i B,
\qquad
\sigma=\pm 1.
\ee
The relation to the Liouville equation is immediate. Indeed, \eqref{eq: self-dual eq} implies
\be
\frac{mE^i}{B}
=
\sigma\frac{\hbar}{2}\,\partial_i\ln B,
\ee
and substituting this into \eqref{eq: Gauss law CS app} gives
\be
\nabla^2\ln B
=
-\sigma\, \frac{2e^2}{\kappa\hbar}\,B.
\ee
For non-negative $B$, the self-dual branch must therefore be chosen with $\sigma=\mathrm{sgn}(\kappa)$.
The radially symmetric family solution to the Liouville equation on the regular region $r>0$ then takes the form\footnote{For $|n|>1$, the density vanishes at $r=0$, so the hydrodynamic variables and the equation for $\ln B$ are understood here only on the regular region $r>0$ where $B>0$. If the equation is extended distributionally to include the origin, $\nabla^2\ln B$ acquires the additional term $4\pi(|n|-1)\delta^{(2)}(\mathbf x)$. We do not consider this distributional extension here.}
\be
B(r)
=
\frac{4|n|^2}{\alpha\,r^2}
\left[
\left(\frac{r_0}{r}\right)^{|n|}
+
\left(\frac{r}{r_0}\right)^{|n|}
\right]^{-2},
\qquad |n|\ge 1,
\ee
where 
\be
\alpha\equiv \frac{e^2}{|\kappa|\hbar}>0.
\ee
The flux of the hydrodynamic gauge field is finite and equal to
\be
\Phi_A\equiv \int d^2x\,B
=
\lim_{r\to\infty}2\pi rA_\theta(r)
=
\frac{4\pi |n|}{\alpha}.
\ee
To get the flux quantization we calculate the global condition \eqref{eq:quant cond in CS}. Since $v^i=\varepsilon^{ij}E_j/B$, the self-dual equation gives
\be
m v_\theta(r)
=
-\sigma\,\frac{\hbar}{2}\,\partial_r\ln B.
\ee
For the above radial profile this becomes
\be
m v_\theta(r)
=
-\sigma\,\frac{\hbar}{r}
\left(
|n|-1-\frac{2|n|u}{1+u}
\right),
\ee
where
\be
u\equiv \left(\frac{r}{r_0}\right)^{2|n|}.
\ee
Choosing radial gauge $A_r=0$, one integrates $B=(1/r)\partial_r(rA_\theta)$ to obtain\footnote{On the punctured plane the general radial solution also permits an
additional flat contribution $C/r$ to $A_\theta$. Here we choose $C=0$, which is the representative used in the
subsequent period calculation.}
\be
A_\theta(r)
=
\frac{1}{r}\int_0^r ds\,sB(s)
=
\frac{2|n|}{\alpha r}\frac{u}{1+u}.
\ee
Using $\sigma=\mathrm{sgn}(\kappa)$, one finds
\be
m v_\theta(r)-\frac{e^2}{\kappa}A_\theta(r)
=
-\sigma\,\frac{\hbar(|n|-1)}{r}.
\ee
On a circle $C_r$ of radius $r$ one obtains
\be
\oint_{C_r}
\left(
 m\mathbf v-\frac{e^2}{\kappa}\mathbf A
\right)\!\cdot d\mathbf l
=
-\sigma\,2\pi\hbar(|n|-1).
\ee
Imposing the global condition \eqref{eq:quant cond in CS} therefore yields
\be
|n|-1\in\mathbb Z.
\ee
Together with the regularity condition on the density profile, this implies $|n|\in\mathbb Z_{>0}$. The integer appearing in the fundamental period condition is thus the phase winding $|n|-1$, while the magnetic flux is labeled by $|n|$, exactly as in \cite{Jackiw:1990tz}.

Finally, since from (\ref{eq:a_mu solution}) we have
\be
b=-\frac{e}{\kappa}B,
\ee
the magnetic flux of the dynamical Chern-Simons field is
\be
\Phi_a\equiv \int d^2x\,b
=
-\frac{e}{\kappa}\Phi_A
=
\pm\frac{4\pi\hbar}{e}|n|,
\ee
where the overall sign is fixed by the chosen self-dual branch.

\section{From the gauge action to spin hydrodynamics}
\label{app:derivation}

We start from the gauge action \eqref{eq:minimal spin-gauge action} and derive its equations of motion in hydrodynamic form. We then take the two-component Schrodinger equation, perform the Madelung decomposition \eqref{eq:spin Madelung}, and show that it produces exactly the same hydrodynamic system. Throughout this appendix we use the sign conventions of Appendix~\ref{app:notation}
\be
a_0=i\hbar\,\xi^\dagger\partial_t\xi,
\qquad
a_i=-\,i\hbar\,\xi^\dagger\partial_i\xi.
\ee

\subsection*{Gauge action \(\rightarrow\) hydrodynamic equations}

The gauge action is
\begin{multline}
I[A_\mu,\xi,\Lambda]
=
\int dt\,d^2x\,
\Bigg[
\frac{m\mathbf E^2}{2B}
-U(B,\nabla B)
-\frac{\hbar^2}{2m}\,
B\,\mathfrak D_i\xi^\dagger\mathfrak D_i\xi
\Bigg]
\\
-\int dt\,d^2x\;
\varepsilon^{\mu\nu\rho}A_\mu\partial_\nu a_\rho
+\int dt\,d^2x\;\Lambda(\xi^\dagger\xi-1),
\label{eq:spin-gauge-action-app}
\end{multline}
where
\be
\mathfrak D_t\xi
=
\partial_t\xi+\frac{i}{\hbar}a_0\xi,
\qquad
\mathfrak D_i\xi
=
\partial_i\xi-\frac{i}{\hbar}a_i\xi,
\label{eq:spin-gauge-covariant-derivatives-app}
\ee
and therefore
\be
\xi^\dagger\mathfrak D_t\xi=0,
\qquad
\xi^\dagger\mathfrak D_i\xi=0.
\ee

As in the spinless case, the gauge field already solves the continuity equation identically through
\be
\rho\equiv B,
\qquad
j^i\equiv \varepsilon^{ij}E_j,
\qquad
j^i=\rho\,v^i.
\label{eq:spin-gauge-hydro-ident-app}
\ee
The Bianchi identity then gives
\be
\partial_t\rho+\partial_i j^i=0,
\qquad
\partial_t\rho+\partial_i(\rho v^i)=0.
\label{eq:spin-gauge-continuity-app}
\ee

The \(A_0\) variation gives
\be
\varepsilon^{ij}\partial_i\!\left(mv_j-a_j\right)=0.
\label{eq:spin-gauge-vorticity-app}
\ee
Hence, on any simply connected patch where \(\rho\neq0\), there exists a phase \(S\) such that
\be
m v_i-a_i=\partial_i S,
\qquad\Longleftrightarrow\qquad
m v_i=\partial_i S+a_i.
\label{eq:spin-gauge-momentum-app}
\ee

Next, varying with respect to \(A_i\) gives
\be
\partial_t\!\left(mv_i-a_i\right)
+
\partial_i\!\Bigg(
-a_0+\frac{m\mathbf v^2}{2}
+V+Q
+\frac{\hbar^2}{2m}\,
\mathfrak D_k\xi^\dagger\mathfrak D_k\xi
\Bigg)
=0,
\label{eq:spin-gauge-euler-app}
\ee
Using \eqref{eq:spin-gauge-momentum-app}, this integrates once to
\be
\partial_t S-a_0
+\frac{m\mathbf v^2}{2}
+V+Q
+\frac{\hbar^2}{2m}\,
\mathfrak D_k\xi^\dagger\mathfrak D_k\xi
=0.
\label{eq:spin-gauge-bernoulli-app}
\ee

It remains to vary the action with respect to \(\xi^\dagger\). For this purpose it is convenient to rewrite the BF term up to a total derivative as
\be
-\varepsilon^{\mu\nu\rho}A_\mu\partial_\nu a_\rho
=
\rho\,a_0-j^i a_i.
\label{eq:spin-gauge-BF-rewrite-app}
\ee
Therefore
\be
\delta_{\xi^\dagger}a_0
=
i\hbar\,\delta\xi^\dagger\partial_t\xi,
\qquad
\delta_{\xi^\dagger}a_i
=
-\,i\hbar\,\delta\xi^\dagger\partial_i\xi,
\ee
and the variation of the BF term becomes
\be
\delta_{\xi^\dagger}\!\left(\rho\,a_0-j^i a_i\right)
=
i\hbar\,\rho\,\delta\xi^\dagger\partial_t\xi
+i\hbar\,j^i\delta\xi^\dagger\partial_i\xi.
\label{eq:spin-gauge-BF-variation-app}
\ee

For the stiffness term we use
\be
\delta_{\xi^\dagger}\!\left(\mathfrak D_i\xi\right)
=
-\frac{i}{\hbar}\,\delta a_i\,\xi
=
-\big(\delta\xi^\dagger\partial_i\xi\big)\xi,
\ee
and
\be
\delta_{\xi^\dagger}\!\left(\mathfrak D_i\xi^\dagger\right)
=
\partial_i\delta\xi^\dagger
+\frac{i}{\hbar}a_i\,\delta\xi^\dagger
+\frac{i}{\hbar}\,\delta a_i\,\xi^\dagger
\\
=
\partial_i\delta\xi^\dagger
+\frac{i}{\hbar}a_i\,\delta\xi^\dagger
+\big(\delta\xi^\dagger\partial_i\xi\big)\xi^\dagger.
\ee
Since
\be
\xi^\dagger\mathfrak D_i\xi=0,
\qquad
(\mathfrak D_i\xi)^\dagger\xi=0,
\ee
the terms proportional to \(\delta\xi^\dagger\partial_i\xi\) drop out, and one finds
\begin{multline}
\delta_{\xi^\dagger}
\left[
-\frac{\hbar^2}{2m}\rho\,
\mathfrak D_i\xi^\dagger\mathfrak D_i\xi
\right]
=
-\frac{\hbar^2}{2m}\rho
\left(
\partial_i\delta\xi^\dagger
+\frac{i}{\hbar}a_i\,\delta\xi^\dagger
\right)\mathfrak D_i\xi
\\
=
\delta\xi^\dagger\,
\frac{\hbar^2}{2m}
\left[
\partial_i\!\big(\rho\,\mathfrak D_i\xi\big)
-\frac{i}{\hbar}\rho\,a_i\,\mathfrak D_i\xi
\right].
\label{eq:spin-gauge-stiffness-variation-app}
\end{multline}

The component of the \(\xi^\dagger\) variation parallel to \(\xi\) is absorbed by \(\Lambda\). Projecting orthogonally with
\be
P_\perp=1-\xi\xi^\dagger
\ee
and using \(j^i=\rho v^i\), the \(\xi\) equation becomes
\be
i\hbar\,\rho\,(\mathfrak D_t+\mathbf v\!\cdot\!\mathfrak D)\xi
=
-\frac{\hbar^2}{2m}\,
P_\perp\!\Big[\partial_i\!\big(\rho\,\mathfrak D_i\xi\big)\Big]
+\frac{i\hbar}{2m}\,\rho\,a_i\,\mathfrak D_i\xi.
\label{eq:spin-gauge-xi-app}
\ee
Therefore the gauge action yields the hydrodynamic system
\eqref{eq:spin-gauge-continuity-app},
\eqref{eq:spin-gauge-momentum-app},
\eqref{eq:spin-gauge-bernoulli-app},
and \eqref{eq:spin-gauge-xi-app}.

\subsection*{Spinor Schrodinger equation \(\rightarrow\)  hydrodynamic equations}

We now start from the free spinor Schrodinger equation
\be
i\hbar\,\partial_t\Psi
=
\left(
-\frac{\hbar^2}{2m}\nabla^2+V
\right)\Psi,
\label{eq:spin-schrodinger-app}
\ee
with
\be
\Psi=\sqrt{\rho}\,e^{iS/\hbar}\,\xi,
\qquad
\xi^\dagger\xi=1.
\label{eq:spin-madelung-app}
\ee
Using
\be
\partial_t\xi
=
\mathfrak D_t\xi-\frac{i}{\hbar}a_0\xi,
\qquad
\partial_i\xi
=
\mathfrak D_i\xi+\frac{i}{\hbar}a_i\xi,
\label{eq:spin-partial-from-covariant-app}
\ee
the derivatives of \(\Psi\) take the form
\be
\partial_t\Psi
=
e^{iS/\hbar}
\Bigg[
\left(
\partial_t\sqrt{\rho}
+\frac{i}{\hbar}\sqrt{\rho}\,(\partial_tS-a_0)
\right)\xi
+\sqrt{\rho}\,\mathfrak D_t\xi
\Bigg],
\label{eq:spin-time-derivative-app}
\ee
and
\be
\partial_i\Psi
=
e^{iS/\hbar}
\Bigg[
\left(
\partial_i\sqrt{\rho}
+\frac{i}{\hbar}\sqrt{\rho}\,mv_i
\right)\xi
+\sqrt{\rho}\,\mathfrak D_i\xi
\Bigg].
\label{eq:spin-space-derivative-app}
\ee
To expand the Laplacian, let
\be
F_i
=
\partial_i\sqrt{\rho}
+\frac{i}{\hbar}\sqrt{\rho}\,(\partial_iS+a_i).
\label{eq:spin-Fi-app}
\ee
Then
\be
\partial_i\Psi
=
e^{iS/\hbar}\Big(F_i\xi+\sqrt{\rho}\,\mathfrak D_i\xi\Big),
\ee
and a direct differentiation gives
\begin{multline}
\nabla^2\Psi
=
e^{iS/\hbar}
\Bigg[
\left(
\partial_iF_i
+\frac{i}{\hbar}(\partial_iS+a_i)F_i
\right)\xi
\\
+\left(
2\partial_i\sqrt{\rho}
+\frac{i}{\hbar}\sqrt{\rho}\,(2\partial_iS+a_i)
\right)\mathfrak D_i\xi
+\sqrt{\rho}\,\partial_i\!\left(\mathfrak D_i\xi\right)
\Bigg].
\label{eq:spin-laplacian-expanded-app}
\end{multline}

The \(\xi\)-parallel projection is extracted with \(\xi^\dagger\). Since \(\xi^\dagger\mathfrak D_i\xi=0\), one has
\be
0=\partial_i\!\left(\xi^\dagger\mathfrak D_i\xi\right)
=
(\partial_i\xi^\dagger)\mathfrak D_i\xi
+\xi^\dagger\partial_i(\mathfrak D_i\xi).
\ee
Using
\be
\partial_i\xi^\dagger
=
\mathfrak D_i\xi^\dagger-\frac{i}{\hbar}a_i\xi^\dagger,
\ee
this becomes
\be
\xi^\dagger\partial_i(\mathfrak D_i\xi)
=
-\mathfrak D_i\xi^\dagger\mathfrak D_i\xi.
\label{eq:spin-project-partialD-app}
\ee
Moreover,
\be
\partial_iF_i
+\frac{i}{\hbar}(\partial_iS+a_i)F_i
=
\nabla^2\sqrt{\rho}
-\frac{\sqrt{\rho}}{\hbar^2}(\nabla S+\mathbf a)^2
+\frac{i}{\hbar\sqrt{\rho}}\,
\partial_i\!\Big(\rho\,(\partial_iS+a_i)\Big).
\label{eq:spin-parallel-coefficient-app}
\ee
Substituting \eqref{eq:spin-time-derivative-app},
\eqref{eq:spin-laplacian-expanded-app},
\eqref{eq:spin-project-partialD-app},
and \eqref{eq:spin-parallel-coefficient-app} into \eqref{eq:spin-schrodinger-app}, the imaginary part of the \(\xi\)-parallel projection yields
\be
\partial_t\rho
+\frac{1}{m}\,
\partial_i\!\Big(\rho\,(\partial_iS+a_i)\Big)
=0.
\label{eq:spin-sch-continuity-app}
\ee
This shows that the transport velocity is
\be
m v_i=\partial_iS+a_i.
\label{eq:spin-sch-velocity-app}
\ee
With this identification the continuity equation becomes
\be
\partial_t\rho+\partial_i(\rho v^i)=0,
\ee
in agreement with \eqref{eq:spin-gauge-continuity-app}.

The real part of the same projection gives
\be
\partial_tS-a_0
+\frac{1}{2m}(\nabla S+\mathbf a)^2
+V+Q
+\frac{\hbar^2}{2m}\,
\mathfrak D_i\xi^\dagger\mathfrak D_i\xi
=0,
\label{eq:spin-sch-bernoulli-app}
\ee
or equivalently,
\be
\partial_tS-a_0
+\frac{m\mathbf v^2}{2}
+V+Q
+\frac{\hbar^2}{2m}\,
\mathfrak D_i\xi^\dagger\mathfrak D_i\xi
=0,
\ee
which is the same equation as \eqref{eq:spin-gauge-bernoulli-app}.

It remains to project orthogonally to \(\xi\). Since \(P_\perp\xi=0\) and \(P_\perp\mathfrak D_i\xi=\mathfrak D_i\xi\), \eqref{eq:spin-laplacian-expanded-app} gives
\be
P_\perp(\nabla^2\Psi)
=
e^{iS/\hbar}
\Bigg[
\left(
2\partial_i\sqrt{\rho}
+\frac{i}{\hbar}\sqrt{\rho}\,(2\partial_iS+a_i)
\right)\mathfrak D_i\xi
\\
+P_\perp\!\Big[\sqrt{\rho}\,\partial_i(\mathfrak D_i\xi)\Big]
\Bigg].
\label{eq:spin-laplacian-orthogonal-app}
\ee
The first and third terms may be reorganized as
\begin{multline}
\left(
2\partial_i\sqrt{\rho}
+\frac{i}{\hbar}\sqrt{\rho}\,(2\partial_iS+a_i)
\right)\mathfrak D_i\xi
+P_\perp\!\Big[\sqrt{\rho}\,\partial_i(\mathfrak D_i\xi)\Big]
\\
=
\frac{1}{\sqrt{\rho}}\,
P_\perp\!\Big[\partial_i\!\big(\rho\,\mathfrak D_i\xi\big)\Big]
+\frac{2i}{\hbar}\sqrt{\rho}\,(\partial_iS+a_i)\mathfrak D_i\xi
-\frac{i}{\hbar}\sqrt{\rho}\,a_i\mathfrak D_i\xi.
\label{eq:spin-laplacian-rewrite-app}
\end{multline}
Using \eqref{eq:spin-sch-velocity-app}, this becomes
\begin{multline}
P_\perp(\nabla^2\Psi)
=
e^{iS/\hbar}
\Bigg[
\frac{1}{\sqrt{\rho}}\,
P_\perp\!\Big[\partial_i\!\big(\rho\,\mathfrak D_i\xi\big)\Big]
+\frac{2im}{\hbar}\sqrt{\rho}\,v_i\mathfrak D_i\xi
-\frac{i}{\hbar}\sqrt{\rho}\,a_i\mathfrak D_i\xi
\Bigg].
\label{eq:spin-laplacian-orthogonal-final-app}
\end{multline}
Now the orthogonal projection of \eqref{eq:spin-schrodinger-app} gives
\be
i\hbar\,\rho\,\mathfrak D_t\xi
=
-\frac{\hbar^2}{2m}\,
P_\perp\!\Big[\partial_i\!\big(\rho\,\mathfrak D_i\xi\big)\Big]
-i\hbar\,\rho\,v_i\mathfrak D_i\xi
+\frac{i\hbar}{2m}\,\rho\,a_i\mathfrak D_i\xi,
\ee
or, after moving the convective term to the left,
\be
i\hbar\,\rho\,(\mathfrak D_t+\mathbf v\!\cdot\!\mathfrak D)\xi
=
-\frac{\hbar^2}{2m}\,
P_\perp\!\Big[\partial_i\!\big(\rho\,\mathfrak D_i\xi\big)\Big]
+\frac{i\hbar}{2m}\,\rho\,a_i\,\mathfrak D_i\xi.
\label{eq:spin-sch-xi-app}
\ee
This is exactly the same spin-transport equation as on the gauge side,
\eqref{eq:spin-gauge-xi-app}.

Therefore the Madelung decomposition of the spinor Schrodinger equation yields precisely the same hydrodynamic equations as the gauge action,
\be
\partial_t\rho+\partial_i(\rho v^i)=0,
\qquad
m v_i=\partial_iS+a_i,
\ee
\be
\partial_tS-a_0
+\frac{m\mathbf v^2}{2}
+V+Q
+\frac{\hbar^2}{2m}\,
\mathfrak D_i\xi^\dagger\mathfrak D_i\xi
=0,
\ee
and
\be
i\hbar\,\rho\,(\mathfrak D_t+\mathbf v\!\cdot\!\mathfrak D)\xi
=
-\frac{\hbar^2}{2m}\,
P_\perp\!\Big[\partial_i\!\big(\rho\,\mathfrak D_i\xi\big)\Big]
+\frac{i\hbar}{2m}\,\rho\,a_i\,\mathfrak D_i\xi.
\ee

\end{document}